\documentclass[final, 12pt, sort&compress]{xelsarticle} 

\RequirePackage[top=2.5cm,bottom=2.5cm,left=2.5cm,right=2.5cm]{geometry}

\def\dyncolwidth{252pt}
\def\ScaleEQU{0.9}

\journal{arXiv}  

\RequirePackage{cj}

\begin{document}

\begin{frontmatter}

    \title{Accelerating off-lattice kinetic Monte Carlo simulations to predict hydrogen vacancy-cluster interactions in \texorpdfstring{$\alpha$}{alpha}--Fe}

    \cortext[cor1]{Corresponding author}

    \author[cam]{C.J. Williams\corref{cor1}}
    \ead{cw648@cam.ac.uk}

    \affiliation[cam]{
        organization={Department of Materials Science and Metallurgy},
        addressline={University of Cambridge},
        city={Cambridge},
        postcode={CB3 0FS},
        country={UK}
    }

    \author[cam,ucl]{E.I. Galindo-Nava\corref{cor1}}
    \ead{e.galindo-nava@ucl.ac.uk}

    \affiliation[ucl]{
        organization={Department of Mechanical Engineering},
        addressline={University College London},
        city={London},
        postcode={WC1E 6BT},
        country={UK}
    }

    \begin{abstract}
        We present an enhanced off-lattice kinetic Monte Carlo (OLKMC) model, based on a new method for tolerant classification of atomistic local-environments that is invariant under Euclidean-transformations and permutations of atoms. Our method ensures that environments within a norm-based tolerance are classified as equivalent. During OLKMC simulations, our method guarantees to elide the maximum number of redundant saddle-point searches in symmetrically equivalent local-environments. Hence, we are able to study the trapping/detrapping of hydrogen from up to five-vacancy clusters and simultaneously the effect hydrogen has on the diffusivity of these clusters. These processes occur at vastly different timescales at room temperature in body-centred cubic iron. We are able to predict the diffusion pathways of clusters/complexes without \emph{a priori} assumptions of their mechanisms, not only reproducing previously reported mechanisms but also discovering new ones for larger complexes. We detail the hydrogen-induced changes in the clusters' diffusion mechanisms and find evidence that, in contrast to mono-vacancies, the introduction of hydrogen to larger clusters can increase their diffusivity. We compare the effective hydrogen diffusivity to Oriani's classical theory of trapping, finding general agreement and some evidence that hydrogen may not always be in equilibrium with traps, when the traps are mobile. Finally, we are able to compute the \emph{trapping atmosphere} of meta-stable states surrounding non-point traps, opening new avenues to better understand and predict hydrogen embrittlement in complex alloys.
    \end{abstract}


\end{frontmatter}

\section{\label{sec:intro}Introduction}

It has been known for over \num{100} years~\cite{JOHNSON1875, Reynolds1875} that the presence of hydrogen (H) in metals -- particularly steels -- can severely reduce ductility, leading to catastrophic failure below the yield-stress. The processes that cause these effects are collectively termed \emph{hydrogen embrittlement} (HE). Despite a century of research, the core mechanisms of HE have yet to be fully understood and are still a topic of active research/debate \cite{Barrera2018}. The difficulty in understanding HE stems from its multi-scale nature; a full description of HE requires understanding of H-adsorption, H-diffusion/transport, and (most crucially) H interaction/influence with/on crystal defects. These processes span many orders of length/time scales, which presents challenges when isolating/connecting the impact of H at the atomistic scale to the macroscopic results.

A breadth of mechanisms for HE have been proposed, most of them revolve around the interactions between H and crystal defects. For a more complete description see \refcite{Dear2017} and \refcite{Barrera2018}. A few of the most prominent mechanisms are: Hydrogen-induced decohesion (HID)~\cite{Pfeil1926,Vehoff1980, Gerberich1991}, Adsorption-induced decohesion (AIDE)~\cite{Lynch1979}, Hydrogen-enhanced localised plasticity (HELP)~\cite{Birnbaum1994,Ferreira1998,Lu2001,Robertson2001,Bak2016} and Hydrogen-enhanced strain-induced vacancy (HESIV) formation~\cite{Nagumo2004, McLellan1997, Nazarov2014}. Many of these mechanisms are supported by bodies of experimental work. As few are orthogonal to each other, it is likely that a full description of HE contains a combination of two/three of these mechanisms (alongside some yet undiscovered).

Theoretical and computational modelling play a crucial role in the study of the Fe-H system due to the inherent difficulty in experimental observations of atomic H~\cite{Kim2019}. The low solubility and high diffusivity~\cite{fukai2005the} of H in body-centred cubic (BCC) iron (Fe), combined with the small `nucleus' and low electron density, make direct experimental observations extremely challenging. Instead, techniques such as thermal desorption analysis (TDA)~\cite{Raina2018}, electro-permeation (EP) experiments~\cite{Raina2017} and atom probe tomography (APT)~\cite{Takahashi2012} are employed. Many of these methods (with the notable exception of APT) are unable to directly investigate H diffusion and trapping behaviour within metals on the atomic scale thus, we must fall back to computation/theory to unravel the atomic mechanisms that cause HE.

Different computational modelling techniques have been used to investigate HE over varied assumptions and time/length scales. On the smallest length-scales, density functional theory (DFT) is used to study H binding sites~\cite{Counts2010,Hayward2013,Nazarov2010} and occasionally combined with molecular dynamic (MD) in \emph{ab initio} MD to study H diffusion at the highest accuracy~\cite{Sanchez2010}. Additionally, work has been done using path-integral MD~\cite{Kimizuka2011, Katzarov2013} to explore H diffusion in iron while incorporating quantum effects, which are known to be important at low temperatures~\cite{Kimizuka2011}. Nevertheless, progress has been made modelling much-larger systems using classical approaches, the most popular of these is MD and its accelerated-variants using semi-empirical potentials. This has enabled the study of H-defect kinetics, such as grain-boundaries~\cite{Wang2016,Teus2016} and dislocations~\cite{Song2012, Tehranchi2016}. Molecular dynamics simulations must resolve atomic vibrations in order to accurately track the dynamics of atom-scale systems. This imposes a significant computational effort as the integration time-step must be of-the-order of these vibrations. Hence, even using today's computers, MD simulation timescales rarely exceed $\bigO{100\si{\us}}$. Continuing to the longest/largest scales, Monte-Carlo (MC)~\cite{Hastings1970} and kinetic Monte-Carlo (KMC)~\cite{Bortz1975} methods overcome this barrier by ignoring the explicit phase-space trajectory, instead focusing on state$\to$state transitions. This can significantly accelerate simulations. However, these methods are confined to discrete representation and require knowledge of mechanisms \emph{a priori}.

Off-lattice\footnote{Also known as: \emph{``adaptive'', ``on-the-fly'', ``self-learning''} and \emph{``self-evolving''}.} kinetic Monte Carlo (OLKMC)~\cite{Henkelman2001}, an extended KMC method, is a general and unbiased tool (discovering mechanisms without any \emph{a priori} input) being successfully applied to study the kinetics of various systems, \eg Fe/Cu/Al, BBC/FCC, disordered systems, extended defects, point defects, etc.~\cite{Xu2009,Pedersen2009,ElMellouhi2008,Lu2016,Restrepo2016}.  Off-lattice KMC allows for the exploration of continuous systems, at previously inaccessible timescales, at atomic fidelity. As such, it is the perfect tool to explore the uncertain and complex mechanisms controlling HE.

In this paper we develop enhancements to the OLKMC method. The motivation for our research is building a general simulation framework capable of modelling the complex interactions between crystal defects and H in iron, into the timescales required to study the mechanisms of HE. Our main contribution is an error-tolerant atomic local-environment (LE) identification/matching process to elide saddle-point searches.

We apply our OLKMC implementation to study the diffusion of vacancy clusters in the presence of H. We demonstrate OLKMC is capable of reaching embrittlement timescales, of-the-order-of seconds, while simultaneously resolving the atomic motion of H-atoms. With OLKMC we can study the atomic mechanisms through which H affects the diffusion of vacancy clusters. These are important first steps towards modelling the more complex H-defect interactions required to gain a full understanding of HE.

\section{\label{sec:olkmc_back}Background: off-lattice kinetic Monte Carlo }

The OLKMC method~\cite{Henkelman2001} operates on a system of $n$ atoms in continuous space and encodes atomic interactions via a potential-energy function:
\begin{align}
    U \colon \mathbb{R}^{3n} \to \mathbb{R}
\end{align}
Local minima (basins) of $U$ are states of the system and are linked together, into a Markov chain, by the mechanisms between them and their corresponding rate constants:
\begin{align}
    \Gamma_{ij} \colon \text{probability per unit time of transition $i\to j$}
\end{align}
The rejection-free $n$-fold way KMC algorithm~\cite{Bortz1975} is used to advance the system forward; with the system in state $i$, the next state $k$ is selected as the solution to:
\begin{align}
    \sum_{j = 1}^{k - 1} \Gamma_{ij} < \rho_1 \sum_{j = 1}^{n} \Gamma_{ij} \leq \sum_{j = 1}^{k} \Gamma_{ij} \label{equ:kmc_choice}
\end{align}
where $\rho_1 \in (0,1]$ is a uniform random number and $j,k \in \{1,2,\ldots,n\}$. The rate-constants model a Poisson process and therefore the time elapsed during a single MC step is~\cite{Bombac2017}:
\begin{align}
    \Delta t = \frac{-\ln \left( \rho_2 \right)}{ \sum_{j = 1}^{n} \Gamma_{ij}} \label{equ:kmc_time}
\end{align}
where $\rho_2 \in (0,1]$ is a second uniform random number. In traditional KMC, these rate constants are typically pre-computed and the catalogue of mechanisms known \textit{a priori}. By applying the harmonic transition state theory (HTST) approximation, the rate constant connecting basins $i\to j$ via the single, first-order, saddle-point~(SP)~$\ddagger$ is described by the Arrhenius equation~\cite{Vineyard1957}:
\begin{align}
    \Gamma^\text{TST}_{ij} = \tilde{\nu}_{ij} e^{-\beta\left( E^\ddagger - E^i \right)} \label{equ:arr}
\end{align}
where $\beta = \frac{1}{k_B T}$, $\tilde{\nu}_{ij}$ is the attempt frequency or Arrhenius prefactor and $E^\ddagger$, $E^i$ are the energies of the system at the SP and state $i$, respectively. Hence, it is possible to build/adapt the atomistic mechanism catalogue on-the-fly by searching $U$ for these SPs.

\subsection{Saddle-point searches}

The process of finding SPs, called the saddle-point search (SPS) procedure, is critical to the efficiency of OLKMC simulations. Minimum-mode following methods~\cite{Cerjan1981} for single ended searches (starting from a single basin) find SPs of the potential energy surface (PES) by climbing from a local minima to an adjacent SP. This is achieved by inverting the component of the force parallel to the minimum eigenmode of the PES. Translating along this force maximises the energy along the minimum-mode and minimises the energy along all other modes hence, converging to a local SP~\cite{Kstner2008}.

Several minimum-mode following algorithms were unified under one mathematical framework in \refcite{Zeng2014} and compared. All investigated methods are bounded in efficiency by the \citeauthor{Lanczos1950} method~\cite{Lanczos1950}. We choose to use the superlinear dimer-method~\cite{Kstner2008}, owing to it requiring fewer force evaluations but converging almost as fast as the \citeauthor{Lanczos1950} method. The superlinear dimer-method contains several optimisations over the original dimer-method~\cite{Henkelman1999} -- notably the improvements of \refcite{Heyden2005}. We discuss a minor modification in \cref{sec:sps}.

\subsection{Saddle-point recycling}

Each SPS requires many hundreds of calls to the force-field and many SPS must be carried out to ensure the completeness of the KMC catalogue. Due to the local nature of mechanisms, most of these SPS are unnecessary. For example, in a section of perfect lattice the LE around each atom is identical hence, the mechanisms that can occur at each atom are identical. Secondly, consider two atoms sufficiently far apart; a local mechanism centred on one will likely not change the LE around the second therefore, its accessible mechanisms remain the same. Finally, many atoms are in LEs differing only by an Euclidean transformation (of the form $\bm{r} \mapsto \bm{Rr} + \bm{c}$, with $\bm{R}$ an orthogonal matrix) hence, their mechanisms are related by the same transformation.

Multiple methods have been developed to reduce the cost of building the KMC catalogue by exploiting this locality. The simplest of these are system-wide methods, which attempt to reuse SPs discovered at the previous step~\cite{Xu2008} however, these do not exploit any relevant symmetries. Due to this inefficiency, they will not be discussed further. Alternatively, local methods seek to classify the LE around each atom in the system. It is then possible to associate mechanisms entirely within a LE. Mechanisms can then be cached and, when an equivalent LE is discovered, instead of launching new SPS, the mechanisms can be reconstructed from the cached information. If the LE classification is suitably invariant, these methods can account for all relevant symmetries hence, the focus shifts to LE classification, of which a number of methods have been proposed.

\paragraph{Space discretisation}Discrete pattern recognition methods for LE classification have been explored~\cite{Shah2012, Nandipati2012} however, these often fail to account for (continuous) symmetries and, because of the discretisation of space, are sensitive to small changes in atomic positions (due to inexact energy minimisations).

\paragraph{Norm-based}Moving toward a tolerant classification, \refcite{Konwar2011} presents a system that stores the LE of an atom at $\bm{r}_i$ as $\Set{\bm{r}_{ij} | r_{ij} < r_\text{env}}$.  Local environments are considered equivalent when each atom in two superimposed LEs have a corresponding atom in the second environment within tolerance $\Delta r_\text{tol}$. This method gracefully allows for error on the positions of atoms in a LE. However, no method is presented for determining this equivalence between arbitrarily permuted/transformed LEs.

\paragraph{Topological} Graph-based topological methods, introduced in \refcite{ElMellouhi2008}, fully exploit the symmetry of LEs. Atoms in the LE are used to draw a graph; atoms become nodes and atoms considered bonded (closer than some distance) are connected with an edge. LEs are equivalent if their graph representations are isomorphic. This is, in general, a problem in its own complexity class GI $\in$ NP which is not known to be in either P or NP-complete~\cite{FORTIN1996}. Fortunately, there exists implementations such as the \texttt{nauty}\footnote{\url{https://pallini.di.uniroma1.it/}} software~\cite{McKay2014} which can solve this problem in polynomial time for many graphs. Although powerful, topological methods rely on a one-to-one correspondence between topology and geometry that may breakdown. Furthermore, they lose the tolerance of norm-based methods.

In \cref{sec:tol}, we introduce our own norm-based LE classification that combine the desirable properties of many of the previous methods.

\subsection{Superbasins and the low-barrier problem}

A common issue encountered during OLKMC simulations is the \emph{low-barrier problem} (LBP)~\cite{ElMellouhi2008, Trochet2018}. This occurs when a collection of basins -- often called a \emph{superbasin} -- are connected by a series of fast mechanisms. It requires many MC steps to escape from a superbasin. As the rate-sum, $\sum_{j = 1}^{n} \Gamma_{ij}$ in \cref{equ:kmc_time}, is very large during this period, the simulated time advances very slowly.

The simplest methods to overcome the LBP effectively combine states connected by fast mechanisms into a single state and ignore all internal superbasin kinetics~\cite{Ramasubramaniam2008} -- this is clearly not exact. Alternatively, TABU-like~\cite{Glover1998} methods that ban recent-transitions have been employed~\cite{Chubynsky2006, ElMellouhi2008}. These have been shown to be thermodynamically sound providing the total number of KMC steps is much greater than the oldest banned transition. Two exact solutions to the LBP are presented in \refcite{Fichthorn2013}; the key insight is the partitioning of states into \emph{transient} and \emph{absorbing} sets, followed by analytically solving the motion inside the transient states. A similar exact solution, the \emph{mean-rate method} (MRM)~\cite{Puchala2010}, has been extended to OLKMC to form the basin auto-constructing MRM (bac-MRM)~\cite{Bland2011}, which constructs superbasins on-the-fly. We discuss minor extensions in \cref{sec:bacmrm}.

\section{\label{sec:method}Methodology}

\subsection{Interatomic potentials}

In order to reach HE timescales -- of-the-order-of seconds -- we employ embedded atom method (EAM) potentials~\cite{Finnis1984}. These are short-range, fast, well tested, semi-empirical models of the potential energy of a collection of atoms. Although they are not without criticism~\cite{Mller2014}, EAM potentials have become well established in the literature, particularly for metallic systems. We use the variation presented and fitted in \refcite{Ramasubramaniam2009}, which generalise the EAM embedding function and are fit to first-principles (DFT) measurements and experimental data. Also, fit to a wide variety of targets, the potentials provide good reproduction of several crystal defect structures. We also include the modifications of \refcite{Song2012}, the introduction of additional H-H repulsion, to reduce the H clustering observed in the original potentials.

\subsection{\label{sec:tol}New invariant and tolerant local-environment classification}

\begin{figure*}[t]
    \centering
    \subfloat[\label{fig:topo_a}Reference point-cloud.]{
        \begin{tikzpicture}[
                scale=\ScaleEQU,
                ref/.style={circle, draw=black , very thick, minimum size=8mm},
                refC/.style={regular polygon,regular polygon sides=4, draw=black , very thick, minimum size=10mm},
                mut/.style={circle, draw=black!20, fill=black!20, very thick, minimum size=8mm},
                mutC/.style={regular polygon,regular polygon sides=4, draw=black!20, fill=black!20, very thick, minimum size=10mm}]

            \node[refC] at (-1,-1) (1){$2$};
            \node[ref] at (-1,1)  (2){$3$};
            \node[ref] at  (1,-1) (4){$5$};
            \node[ref] at  (1,1)  (3){$4$};

            \node[ref] at  (0,0)  (0){$1$};

            \draw[-stealth] (0) -- (1) ;
            \draw[-stealth] (1) -- (2) ;
            \draw[-stealth] (2) -- (3) ;
            \draw[-stealth] (3) -- (4) ;
            \draw[-stealth] (4) -- (0) ;

        \end{tikzpicture}
    } \hfill
    \subfloat[\label{fig:topo_b}Unclassified point-cloud.]{
        \begin{tikzpicture}[
                scale=\ScaleEQU,
                ref/.style={circle, draw=black , very thick, minimum size=8mm},
                refC/.style={regular polygon,regular polygon sides=4, draw=black , very thick, minimum size=10mm},
                mut/.style={circle, draw=black!20, fill=black!20, very thick, minimum size=8mm},
                mutC/.style={regular polygon,regular polygon sides=4, draw=black!20, fill=black!20, very thick, minimum size=10mm}]

            \node[mut] at (-0.50262794, -1.32942286) (1){$2$};
            \node[mut] at (-1.4160254 ,  0.45262794) (2){$3$};
            \node[mut] at ( 1.37272413, -0.47762794) (3){$4$};
            \node[mutC] at ( 0.3160254 ,  1.45262794) (4){$5$};

            \draw[-stealth] (2) -- (3) ;

            \node[mut] at  (0.1,0)  (0) {$1$};

            \draw[-stealth] (0) -- (1) ;
            \draw[-stealth] (1) -- (2) ;
            \draw[-stealth] (3) -- (4) ;
            \draw[-stealth] (4) -- (0) ;

        \end{tikzpicture}
    } \hfill
    \subfloat[\label{fig:topo_c}Permutation of labels in \cref{fig:topo_b}]{
        \begin{tikzpicture}[
                scale=\ScaleEQU,
                ref/.style={circle, draw=black , very thick, minimum size=8mm},
                refC/.style={regular polygon,regular polygon sides=4, draw=black , very thick, minimum size=10mm},
                mut/.style={circle, draw=black!20, fill=black!20, very thick, minimum size=8mm},
                mutC/.style={regular polygon,regular polygon sides=4, draw=black!20, fill=black!20, very thick, minimum size=10mm}]

            \node[mut] at (-0.50262794, -1.32942286) (3){$4$};
            \node[mut] at (-1.4160254 ,  0.45262794) (4){$5$};
            \node[mut] at ( 1.37272413, -0.47762794) (2){$3$};
            \node[mutC] at ( 0.3160254 ,  1.45262794) (1){$2$};
            \node[mut] at ( 0.1       ,  0         ) (0){$1$};

            \draw[-stealth] (0) -- (1) ;
            \draw[-stealth] (1) -- (2) ;
            \draw[-stealth] (2) -- (3) ;
            \draw[-stealth] (3) -- (4) ;
            \draw[-stealth] (4) -- (0) ;

        \end{tikzpicture}
    } \hfill
    \subfloat[\label{fig:topo_d}Rotation of \cref{fig:topo_c}]{
        \begin{tikzpicture}[
                scale=\ScaleEQU,
                ref/.style={circle, draw=black , very thick, minimum size=8mm},
                refC/.style={regular polygon,regular polygon sides=4, draw=black , very thick, minimum size=10mm},
                mut/.style={circle, draw=black!20, fill=black!20, very thick, minimum size=8mm},
                mutC/.style={regular polygon,regular polygon sides=4, draw=black!20, fill=black!20, very thick, minimum size=10mm}]

            \node[mutC] at    (-1,-1.1) (1){$2$};
            \node[mut] at (-0.95,1.1)  (2){$3$};
            \node[mut] at     (1,-1.1) (4){$5$};
            \node[mut] at   (1.1,0.9)  (3){$4$};
            \node[mut] at  (-0.1,0)    (0){$1$};

            \draw[-stealth] (0) -- (1) ;
            \draw[-stealth] (1) -- (2) ;
            \draw[-stealth] (2) -- (3) ;
            \draw[-stealth] (3) -- (4) ;
            \draw[-stealth] (4) -- (0) ;

        \end{tikzpicture}
    } \hfill
    \subfloat[\label{fig:topo_e}\cref{fig:topo_a} over \cref{fig:topo_d}.]{
        \begin{tikzpicture}[
                scale=\ScaleEQU,
                ref/.style={circle, draw=black , very thick, minimum size=8mm},
                refC/.style={regular polygon,regular polygon sides=4, draw=black , very thick, minimum size=10mm},
                mut/.style={circle, draw=black!20, fill=black!20, very thick, minimum size=8mm},
                mutC/.style={regular polygon,regular polygon sides=4, draw=black!20, fill=black!20, very thick, minimum size=10mm}]

            \node[mutC] at    (-1,-1.1)  {};
            \node[mut] at (-0.95,1.1)   {};
            \node[mut] at     (1,-1.1)  {};
            \node[mut] at   (1.1,0.9)   {};
            \node[mut] at  (-0.1,0)     {};

            \node[refC] at (-1,-1) {};
            \node[ref] at (-1,1)  {};
            \node[ref] at  (1,-1) {};
            \node[ref] at  (1,1)  {};
            \node[ref] at  (0,0)  {};

            \node at     (-1, -1.05) {$2$};
            \node at (-0.975,1.05)   {$3$};
            \node at      (1,-1.05)  {$5$};
            \node at   (1.05,0.95)   {$4$};
            \node at  (-0.05,0)      {$1$};

        \end{tikzpicture}
    }

    \caption{Demonstration of the norm-based equivalence between reference point-cloud \cref{fig:topo_a} (opaque shapes) and unclassified point-cloud \cref{fig:topo_b} (grey shapes) via a permutation of the labels and a rigid-body rotation (transformation). Shape (square/circle) denotes the colour of each point  and arrows act as a guide to the eyes for the permutation. In \cref{fig:topo_e} we see all the points/atoms are close enough that the LEs can be considered equivalent.}
    \label{fig:topo}
\end{figure*}
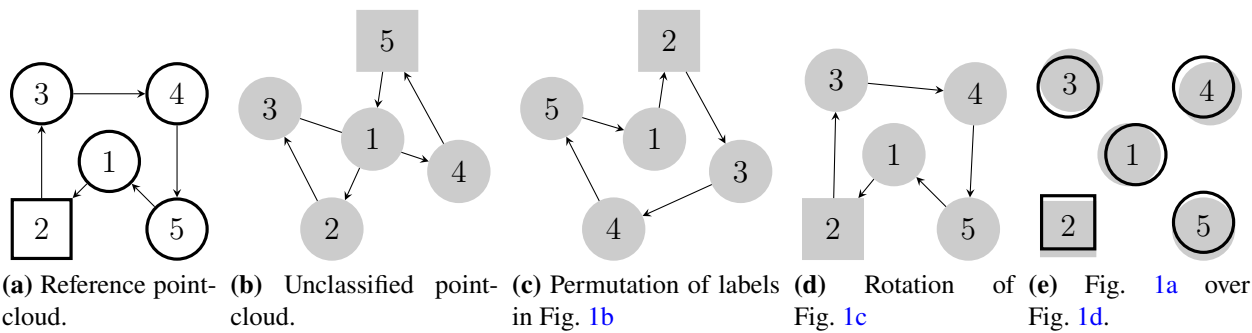

In previous work~\cite{Williams2020}, we adopted a topological classification methodology however, this relied on the aforementioned one-to-one correspondence between topology and geometry. We found this correspondence to break-down in the Fe-H system due to the small size of the H-atom and small displacements during mechanisms. We tried to overcome this problem by allowing the bonding distance to vary with the species and colouring each atom as the pair formed from the atoms' atomic number and the local sum:
\begin{align}
    \left\lfloor c \sum_{j} G_{ji}  r_{ij}  \right\rfloor
\end{align}
with $c$ a problem-dependant scaling constant and $G_{ij}$ elements of the adjacency matrix. This encodes much more of the information contained within $\Set{r_{ij}}$ into the coloured graph. Unfortunately, with the above modifications, infinitesimal perturbations in position are more likely to result in many keys representing the same geometry. Finally, there is no quantitative/qualitative link between topological keys and the similarity of LEs.

\subsubsection{Norm-based definition of equivalence}

We require a notion of equivalence between LEs that is invariant under:
\begin{enumerate}[noitemsep]
    \item Infinitesimal-perturbations of atomic positions.
    \item Permutation of identical atoms.
    \item Euclidean transformations of the group of atoms.
\end{enumerate}
In order to overcome the aforementioned difficulties when dealing with the small errors in the atomic positions we move away from the graph-based representation of the atoms. Instead, we represent atoms as coloured points (\num{2}-tuples):
\begin{align}
    \left(\bm{p} \in \mathbb{R}^3, p \in \mathbb{Z} \right)
\end{align}
and a LE centred on the point $\left(\bm{p}_1, p_1\right)$ as the point cloud:
\begin{align}
    P = \Set{\left(\bm{p}_1, p_1\right), \ldots, \left(\bm{p}_{n}, p_{n}\right)}
\end{align}
where (without loss of generality) we set the centroid of the points in $P$ to the origin and $\norm{\bm{p}_1 - \bm{p}_i} < r_\text{env}$ for all $i$ with $r_\text{env}$ the radius of the LE. The question of determining if two LEs, $P$ and $Q$ (of the same size), are \emph{equivalent} is the same as asking if there exists a transformation matrix $\bm{O}$ and permutation $\pi$ such that:
\begin{align}
    \sum_{i = 1}^n \norm{\bm{p}_i - \bm{Oq}_{\pi\left(i\right)}}^2 \le \delta^2
    \quad \text{and} \quad
    p_i = q_{\pi\left(i\right)} \label{equ:eqiv}
\end{align}
subject to the constraints:
\begin{align}
    \bm{OO}\tran = \bm{O}\tran\bm{O} = I \quad \text{and} \quad \pi\left(1\right) = 1 \label{equ:constraint}
\end{align}
where $\delta$ is the maximum $\ell_2$ norm or \emph{distance} between the point-cloud as well as the maximum inter-point separation:
\begin{align}
    \Delta_i = \norm{\bm{p}_i - \bm{Oq}_{\pi\left(i\right)}}
\end{align}
The choice of $\delta$ controls how similar two LEs must be before they are considered equivalent. This equivalence is represented graphically in \cref{fig:topo}.

By design, relabelling a pair of identical points will always result in an equivalent environment. A desirable property, is to ensuring this relabelling results in a corresponding change in $\pi$ (instead of being absorbed into $\delta^2$) hence, for a useful definition of equivalence, we require:
\begin{align}
    \delta < r_\text{min} \label{equ:den}
\end{align}
where $r_\text{min}$ is the minimum intra-point separation: $r_{ij} = \norm{\bm{r}_i - \bm{r}_j}$. This ensures a consistent correspondence between points in equivalent LEs.

Construction of a point-cloud centred on a point (by selecting points within $r_\text{env}$ from some larger set), is not fully tolerant of point perturbations near the edge of the LE, which could move atoms into/out of the LE. This is acceptable as the LEs will be used for mechanism reconstruction which requires a \num{1}-to-\num{1} correspondence between points. Otherwise, unbalanced equivalence is a natural extension, by simply adding the square of the distance between unmatched points and the boundary of the LE to \cref{equ:eqiv}.

\subsubsection{Connection to the potential energy}

An inequality on $\delta$ can be established by Taylor-expanding the potential energy, $U$, about a converged extrema:
\begin{align}
    \Delta U & \approx \Delta \bm{x}\tran \cancelto{\scriptstyle{\bm{0}}}{\bm{\nabla} U}   + \frac{1}{2}\Delta\bm{x}\tran\bm{H}\Delta\bm{x} \nonumber \\
             & \approx \frac{1}{2}\Delta\bm{x}\tran\bm{Q\Lambda Q}\tran \Delta\bm{x} \label{equ:rtol_hessian}
\end{align}
where we have applied the eigendecomposition~\cite[p.~80]{bishop2006} to the real symmetric Hessian, $\bm{H}$, forming $\bm{\Lambda}$ the diagonal matrix of eigenvalues and $\bm{Q}$, the orthogonal matrix of eigenvectors. Noting an orthogonal transformation does not change the magnitude of a vector. A (weak) upper-bound on $\Delta U$ near a minima can be constructed from \cref{equ:rtol_hessian}:
\begin{align}
    \Delta U & \le \frac{\lambda_\text{max}}{2}\norm{\bm{Q}\tran\Delta\bm{x}}^2 \nonumber \\
             & \le \frac{\lambda_\text{max}}{2}\norm{\Delta\bm{x}}^2            \nonumber \\
             & \le \frac{\lambda_\text{max}}{2} \delta^2 \label{equ:rtol_bound}
\end{align}
where $\lambda_\text{max}$ is the maximum eigenvalue of $\bm{H}$ and the third line follows from \cref{equ:eqiv}. If two LEs are equivalent and $\delta$ is small enough (such that the mechanisms are transferable), then the energy barrier(s) of the reconstructed mechanism(s) should be of-the-order-of $\Delta U$ off the true energy barrier(s). Ultimately, choosing a smaller value of $\delta$ increases the accuracy of the simulation, at the expense of increasing the number of SPS required. Hence, the largest value of $\delta$ that ensures \cref{equ:den} and that $\Delta U$ is much less than the minimum relevant energy-barrier should be chosen.

\subsubsection{Point-cloud registration}

Simultaneously determining the orthogonal transformation and permutation that minimises the $\ell_2$ norm between LEs is a variation of the well-studied rigid point-cloud registration problem~\cite{Zhu2019, Maiseli2017}. In general, this is not possible at the speeds required by OLKMC. However, we are only interested in finding a specific permutation/transformation that satisfies \cref{equ:eqiv} and \cref{equ:constraint}. Here we present a greedy method, that leverages the problem-specific distribution of points.

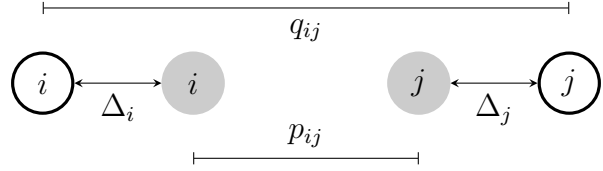
\begin{figure}[t]
    \centering
    \begin{tikzpicture}[
            scale=1,
            ref/.style={circle, draw=black , very thick, minimum size=8mm},
            refC/.style={regular polygon,regular polygon sides=4, draw=black , very thick, minimum size=10mm},
            mut/.style={circle, draw=black!20, fill=black!20, very thick, minimum size=8mm},
            mutC/.style={regular polygon,regular polygon sides=4, draw=black!20, fill=black!20, very thick, minimum size=10mm}]

        \node[mut] at (0, 0) (1){$i$};
        \node[mut] at (3, 0) (2){$j$};

        \node[ref] at (-2, 0) (11){$i$};
        \node[ref] at ( 5, 0) (22){$j$};

        \draw[stealth-stealth,] (1) -- (11) node[midway,anchor=north]{$\Delta_{i}$};
        \draw[stealth-stealth,] (2) -- (22) node[midway,anchor=north]{$\Delta_{j}$};

        \draw[|-|] (-2,  1) -- (5,  1) node[midway,anchor=north]{$q_{ij}$};
        \draw[|-|] ( 0, -1) -- (3, -1) node[midway,anchor=south]{$p_{ij}$};

    \end{tikzpicture}
    \caption{\label{fig:ips}Diagram showing the orientation of two pairs of points $i$ and $j$ in point-clouds $P$ (dark grey) and $Q$ (white) that maximises $\Delta^2_{i} + \Delta^2_{j}$, the sum of the square inter-point separations, under the constraint $\Delta^2_{i} + \Delta^2_{j} \le \delta^2$ from \cref{equ:eqiv}.}
\end{figure}

To find the minimising permutation, we first realise a relationship between the inter-point separations and intra-point separations $p_{ij} = \norm{\bm{p}_i - \bm{p}_j}$ and $q_{ij} = \norm{\bm{q}_i - \bm{q}_j}$ of the respective point-clouds $P$ and $Q$. Studying \cref{fig:ips} we see:
\begin{align}
    \abs{p_{ij} - q_{ij}} & \le \Delta_i + \Delta_j
\end{align}
Then maximising $\Delta_i + \Delta_j$, subject to the constraint from \cref{equ:eqiv}, we find our intra-point tolerance criterion:
\begin{align}
    \abs{p_{ij} - q_{ij}} & \le \sqrt{2} \delta \label{equ:intra}
\end{align}
which can be used to match pairs of points in $Q$ to $P$ by recursively ordering $Q$. At each recursion we search for a point that ensures \cref{equ:intra} holds for all previously ordered points in $Q$. As this method only requires the intra-point separations in $P$ and $Q$ (which are invariant under Euclidean transformations of $P$ and $Q$), this method can match the order of the points in LEs that are related by arbitrary rotations/reflections \emph{before} solving for the rotation/reflection.

Once the order of the points in $P$ and $Q$ match we must solve:
\begin{align}
    \min_{\bm{O} \in M_{3,3}(\mathbb{R})}{\sum_{i=1}^n \norm{\bm{p}_i - \bm{O} \bm{q}_i}^2}
    \quad \text{s.t.}\quad
    \bm{OO}\tran = \bm{O}\tran\bm{O} = \bm{I} \label{equ:opp}
\end{align}
this is equivalent to the \emph{orthogonal Procrustes problem}~\cite{Schnemann1966} which can be efficiently solved using the singular value decomposition (SVD) see \cref{app:opp}.

\begin{algorithm}[htb]
    \caption{\label{alg:permute_onto}Function \textsc{greedy\_perm} and its subroutines attempt to permute elements of $Q$ such that it is equivalant to $P$ (\cref{equ:eqiv} holds);  returns \texttt{True} if $P$ and $Q$ are equivalent otherwise \texttt{False}.}
    \begin{algorithmic}
        \onehalfspacing

        \Require$P$ and $Q$ contain the same number of points, $n > 1$.
        \Function{greedy\_perm}{$P$, $Q$}
        \State \Return \Call{\_recur}{$P$, $Q$, $2$}
        \EndFunction
        \Statex

        \Function{\_recur}{$P$, $Q$, $i$}
        \If{$i > n$}
        \State $\bm{O} \gets \Call{rotor\_onto}{P, Q}$ \Comment{See \cref{app:opp}}
        \State $\Delta^2 \gets \sum_{i=1}^n \norm{\bm{p}_i - \bm{Oq}_{i}}^2 $
        \State \Return $\Delta^2 \le \delta^2$
        \EndIf
        \For{$j \gets i, \ldots, n$}
        \If{$p_i = q_j$}
        \State Swap points $i$ and $j$ in $Q$
        \If{\Call{\_match}{\texttt{P}, \texttt{Q}, $i$}}
        \If{\Call{\_recur}{\texttt{P}, \texttt{Q}, $i + 1$}}
        \State \Return \texttt{True}
        \EndIf
        \EndIf
        \State Swap points $i$ and $j$ in $Q$
        \EndIf
        \EndFor
        \State \Return \texttt{False}
        \EndFunction
        \Statex

        \Function{\_match}{$P$, $Q$, $i$}
        \For{$j \gets 1, \ldots, i - 1$}
        \If{$\abs{p_{ij} - q_{ij}} > \sqrt{2} \delta$}
        \State \Return \texttt{False}
        \EndIf
        \EndFor
        \State \Return \texttt{True}
        \EndFunction
    \end{algorithmic}
\end{algorithm}

The full permutation/ordering and equivalence-testing method is detailed in \cref{alg:permute_onto}; once the algorithm has matched the orders and colours it checks if the permutation satisfies \cref{equ:eqiv}. This is required as there may be degenerate permutations satisfying \cref{equ:intra} $\forall i,j$ but not satisfying \cref{equ:eqiv}. A key consideration for the usefulness of \cref{alg:permute_onto} is its complexity; for two randomly permuted point-clouds, we show in \cref{sec:complexity} (under moderate assumptions) provided:
\begin{align}
    \delta \lesssim \frac{2}{5} r_\text{min} \label{equ:den2}
\end{align}
the average-case time complexity is \bigO{n^2}.

\subsubsection{Choosing \texorpdfstring{$\delta$}{delta} values}

Typically in the Fe-H system (using the perfect lattice as an order of magnitude estimation) $\lambda_\text{max} \approx 10\si{\eV\per\angstrom\squared}$. Therefore, according to \cref{equ:rtol_bound}, we choose $\delta = 0.01$\si{\angstrom} resulting in an energy tolerance of approximately $\Delta U \le \num{5e-4}$\si{\eV}. In practice we expect $\Delta U \ll \num{5e-4}$\si{\eV} as \cref{equ:rtol_bound} assumes $\Delta \bm{x}$ is parallel to the largest eigenvector of $\bm{H}$ which is unlikely. We see $\Delta U$ is much less than the energy barrier for H diffusion, around $\num{5e-2}$\si{\eV}, typically the fastest mechanism in the Fe-H system.

The choice of $\delta$ is continuously validated during a simulation. If $\delta$ is too large then, following a mechanism reconstruction, a relaxation of the lattice will result in a large energy change. If/when this is detected $\delta$ can be adjusted. Conversely, if no such energy changes are detected $\delta$ can be increased to try and increase the performance of the simulation. Furthermore, if we encounter a local environment that breaks \cref{equ:den} or \cref{equ:den2} then $\delta$ can be reduced.

\subsubsection{Heuristics}

With the algorithms detailed thus far, a catalogue could be built that satisfies our requirements for LE classification however, although we ensure the condition of \cref{equ:den2}, a call to \cref{alg:permute_onto} still takes $\approx 10\si{\micro\second}$ with a LE containing \num{65} atoms. Hence, as we may call \cref{alg:permute_onto} for every LE in the catalogue when encountering a new LE, this becomes prohibitively expensive. To reduce the search-space we partition the catalogue into sub-catalogues each indexed by a key, $k$:
\begin{align}
    k \colon P \to \left(p_1, \Set{n^\alpha}\right)
\end{align}
with $n^\alpha$ the number of points in $P$ of the $\alpha^\text{th}$ colour. Due to its discrete nature, $k$ can be used as the key to a hash-table (or other suitable key--value store) enabling \bigO{1} look-up of the sub-catalogues. The sub-catalogues may still become very large, especially in systems where all points are the same colour. As a simulation progresses, we can sort the order of the LEs in each sub-catalogue by its occurrence count. This significantly decreases the look-up time for a typical LE as many systems have most points in the same LE and only a small number of "active" points (e.g near defects) in rare LEs.

To further accelerate searches of the sub-catalogues we introduces a second discriminator, the \emph{fingerprint} $f$, a collection of sorted sets/lists:
\begin{align}
    f \colon P \to \Set{\Set{p_{1i} \mid i > 1}^{\alpha}_\le, \Set{p_{ij} \mid i > 1,j > i}^{\alpha,\, \beta \le \alpha}_\le}
\end{align}
where $p_{ij}$ denotes the intra-point distances between points $i$ and $j$ and the superscripts $\alpha$, $\beta$  indicate the colour of the points in the point pair. For example, in the H-Fe system there are two possible point colours hence, $f$ contains five ordered lists. Two each containing the intra-point distances between points of a particular colour and the central point; a further three lists containing the intra-point distances between pairs of atoms coloured H-H, Fe-H/H-Fe, and Fe-Fe. By construction, $f$ is invariant under Euclidean transformations and permutations of the points in $P$. Two fingerprints can be compared for equivalence as follows:
\begin{algorithm}[H]
    \caption{Compare two fingerprints for equivalence under \cref{equ:eqiv} subject to the constraints of \cref{equ:constraint}.}
    \begin{algorithmic}
        \onehalfspacing
        \Require $P$ and $Q$ have matching keys.
        \Function{equiv}{$f_P$, $f_Q$}
        \For{each pair of ordered lists \texttt{p[]}, \texttt{q[]} in $f_P$, $f_Q$}
        \For{each pair of elements $p$, $q$ in \texttt{p[]}, \texttt{q[]}}
        \If{$\abs{p - q} > \sqrt{2}\delta$}
        \State \Return \texttt{False}
        \EndIf
        \EndFor
        \EndFor
        \State \Return \texttt{True}
        \EndFunction
    \end{algorithmic}
\end{algorithm}
\noindent By construction, it is necessary, but not sufficient, for two LEs fingerprints to be equivalent for \cref{alg:permute_onto} to return \texttt{True}. In practice, equivalence of fingerprints is a very strong pre-conditioner for \cref{alg:permute_onto}. As comparison of fingerprints is orders of magnitude faster (typically taking tens of nanoseconds), this substantially accelerates searching the sub-catalogues.

\subsubsection{Searching a catalogue of LEs}

The full method for classifying a LE, represented by the point-cloud $Q$, and reconstructing the mechanisms discovered by previous SPS at an equivalent LE proceeds as follows:
\begin{algorithm}[H]
    \caption{Search a catalogue of reference LEs for an equivalent LE.}
    \begin{algorithmic}
        \onehalfspacing

        \Function{search\_catalogue}{$Q$}
        \State $k \gets $ the key of $Q$
        \State $s \gets $ the sub-catalogue corresponding to $k$ in the catalogue

        \For{each $P$ in $s$}
        \If{\Call{equiv}{$P$, $Q$} \textbf{and} \Call{greedy\_perm}{$P$, $Q$}}

        \parState{\Return~P \quad an equivalent LE, whose mechanisms can be reconstructed onto $Q$ by multiplying their atomic displacement-vectors by $\bm{O}\tran$ (computed during \textsc{greedy\_perm})}

        \EndIf
        \EndFor

        \State \Return \texttt{NULL}
        \EndFunction

    \end{algorithmic}
\end{algorithm}
\noindent If no match is found then $Q$ represents a new LE; append $Q$ to the sub-catalogue and launch SPS centred on the LE in order to discover any mechanisms associated with it.

\subsection{\label{sec:sps}Saddle-point searches}

In our implementation, we diverge slightly from the original formulation of the superlinear dimer method~\cite{Kstner2008} during the dimer translation step. We still use the L-BFGS algorithm~\cite{Liu1989, Nocedal1980} for determining the translation direction and step size but, introduce a trust-radius based approach to limit the step-size. The maximum step size, $s_\text{trust}$, is scaled according to the success of the previous steps; the projection of the effective gradient on the search direction is calculated after a step:
\begin{align}
    P = -\bm{F}_\text{eff}\tran \bm{p} \label{equ:proj}
\end{align}
where $\bm{p}$ is the approximate Newton step, computed using the L-BFGS method, and $\bm{F}_\text{eff}$ is the effective force acting on the dimer. An ideal step length would have $P = 0$. Hence, we increase $s_\text{trust}$ when $P < -P_\text{tol}$ and decrease $s_\text{trust}$ when $P > P_\text{tol}$. Additionally, we bound $s_\text{trust}$ such that $s_\text{min} < s_\text{trust} < s_\text{max}$.

\subsection{Kinetic prefactors}

Most OLKMC implementations apply the constant pre-factor approximation, $\tilde{\nu}_{ij} = v$, to \cref{equ:arr}~\cite{ElMellouhi2008, Trochet2018} however, \refcite{Chill2014} identifies large variations in $\tilde{\nu}_{ij}$ during Al adatom diffusion events. This is to be expected when dealing with heterogeneous systems and evidence to start calculating $\tilde{\nu}_{ij}$, rather than relying on constant approximation. Applying HTST and assuming the PES is quadratic near the SP, we can calculate $\tilde{\nu}_{ij}$ for each mechanism~\cite{Vineyard1957, Hnggi1990}:
\begin{align}
    \tilde{\nu}_{ij} = \frac{\prod_{k=1}^{N} \nu^i_k}{\prod_{k=1}^{N-1} \nu^\ddagger_k}
\end{align}
where $\nu^\ddagger_k$, $\nu^i_k$ are the real normal-mode frequencies at the SP and state $i$ respectively. This requires computing the full mass-weighted Hessian for each LE. This can be done most efficiently analytically; the procedure is described in the supplementary material.

\subsection{\label{sec:bacmrm}Superbasin caching}

We elect to further extend the bac-MRM to incorporate superbasin caching. We follow \refcite{Bland2011} however, when a superbasin is exited, instead of discarding the superbasin, it is stored into a buffer. The implementation then checks if this new "exit" state is in any of the cached superbasins, if so the corresponding superbasin is loaded from the cache. This is particularly critical as it avoids re-exploring a superbasin that is re-entered immediately after it has been exited. We also dynamically set the tolerance (which the forward and reverse barriers of a mechanism must be less than) for exit state classification. This is achieved by lowering the tolerance when a superbasin gets too large and increasing it when the number of superbasins in the cache exceeds some user-defined threshold.

\subsection{Measuring diffusivity}

The diffusion coefficient of a collection of atoms over a time $t$, can be extracted from the mean-squared displacement (MSD):
\begin{align}
    \langle R^2 \rangle = \frac{1}{N} \sum_\alpha^N \norm{\bm{r}^\alpha_{t=0} - \bm{r}^\alpha_{t}}^2
\end{align}
of the atoms and the Einstein equation~\cite{Lv2018}:
\begin{align}
    D = \frac{\langle R^2 \rangle}{6 t}
\end{align}
As we use a single H atom throughout our simulations, we compute the diffusivity of H by dividing the trajectory into intervals and averaging the diffusivity computed in each interval~\cite{Pascuet2011, Messina2015}. Due to the complexity of explicitly tracking the position of vacancies during a simulation, we use the MSD of the Fe atoms to calculate the simulated diffusivity, $D_\text{sim}$, and effective diffusivity, $D_\text{eff}$ of individual defects. These are related to the MSD via~\cite{EcheverriRestrepo2020}:
\begin{align}
    D_\text{sim} = \frac{\langle R^2 \rangle_\text{Fe}}{6t} \quad \text{and} \quad D_\text{eff} = \frac{D_\text{sim} }{x_d} \label{equ:msd}
\end{align}
with $x_d$ the defect concentration. This has the additional benefit of averaging over many Fe atoms.

\section{\label{sec:rd}Results and discussion}

\subsection{\label{sec:cluster}Vacancy cluster diffusion}

We construct a vacancy-cluster, $\text{V}_{n}$, in a (otherwise perfect) $6^3$ unit-cell BCC supercell.  After a series of mechanisms, the cluster is moved to an equivalent state (just rotated/translated). In combination with the periodic boundary conditions, our norm-based caching recognises this symmetry, reconstructs saddle-points and elides new searches. Hence, SP searches were only required during the initial \emph{learning} phase of the simulation.

\begin{figure}
    \centering
    \includegraphics[width=\dyncolwidth]{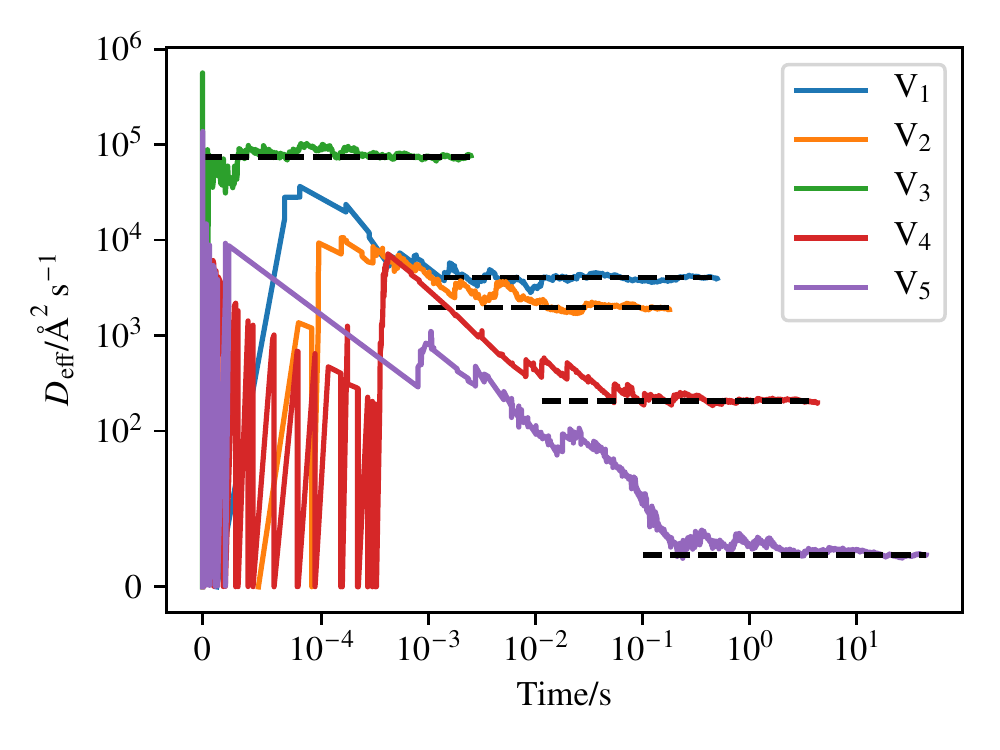}
    \caption{\label{fig:cluster_diff}Vacancy clusters diffusing in a perfect $6^3$ unit-cell supercell at $300$\si{\kelvin}, dashed lines are fit to a constant.}
\end{figure}

\begin{table}[b]
    \centering

    \caption{\label{tab:cluster_diff}Summary of vacancy-cluster diffusion results in the $\alpha$--Fe lattice at \num{300}\si{\kelvin}. All diffusivities have a fractional error less than one part in one hundred. Quoted kinetic pre-factors for multi-step mechanisms is that of the highest barrier step.}

    \renewcommand{\arraystretch}{1.3}

    \begin{tabular*}{\columnwidth}{@{}l@{\extracolsep{\fill}}lll@{}}
        \toprule
        Cluster      & $\Delta E$/\si{\eV} & $v$/$10^{13}$\si{\Hz} & $D_\text{eff}$/\si{\meter\squared\per\second} \\
        \midrule
        $\text{V}_1$ & $0.64(9)$           & \num{7.44}            & \num{4.05e-17}                                \\
        $\text{V}_2$ & $0.65(1)$           & \num{10.4}            & \num{2.07e-17}                                \\
        $\text{V}_3$ & $0.48(2)$           & \num{5.22}            & \num{7.40e-16}                                \\
        $\text{V}_4$ & $0.73(4)$           & \num{3.41}            & \num{2.05e-18}                                \\
        $\text{V}_5$ & $0.77(3)$           & \num{3.12}            & \num{9.01e-20}                                \\
        \bottomrule
    \end{tabular*}

\end{table}

\begin{figure*}[t]
    \centering

    \subfloat[\label{fig:v1}$\text{V}_1$]{
        \includegraphics[width=0.32\textwidth]{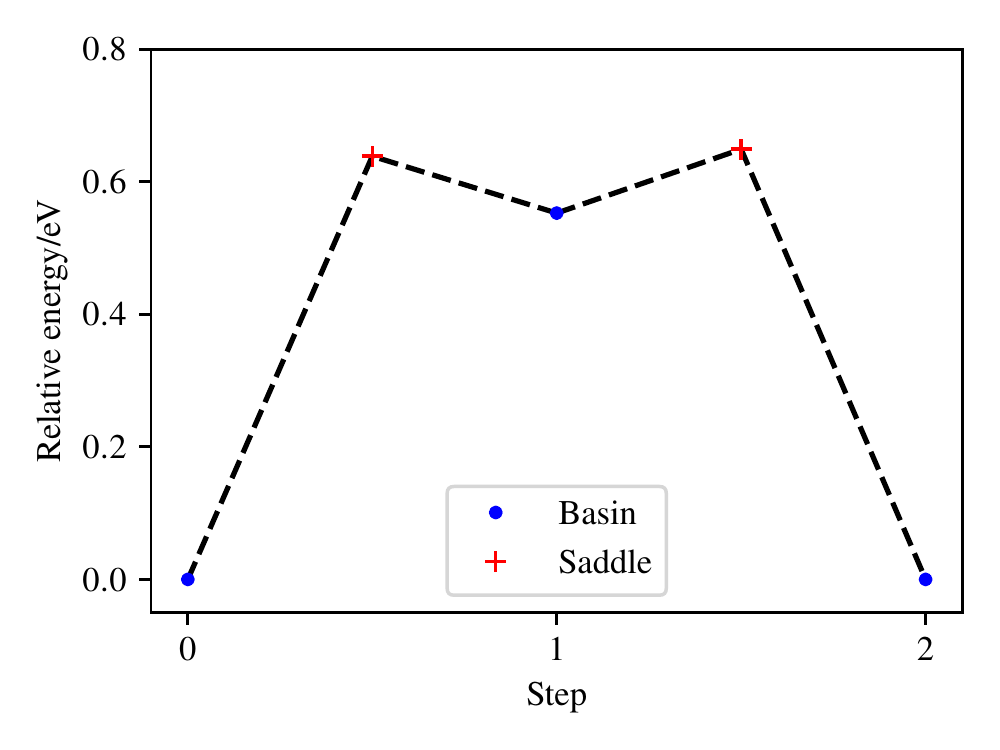}
    } \hfill
    \subfloat[\label{fig:v2_lo}$\text{V}_2$ via $4$\textsuperscript{th}~NN]{
        \includegraphics[width=0.32\textwidth]{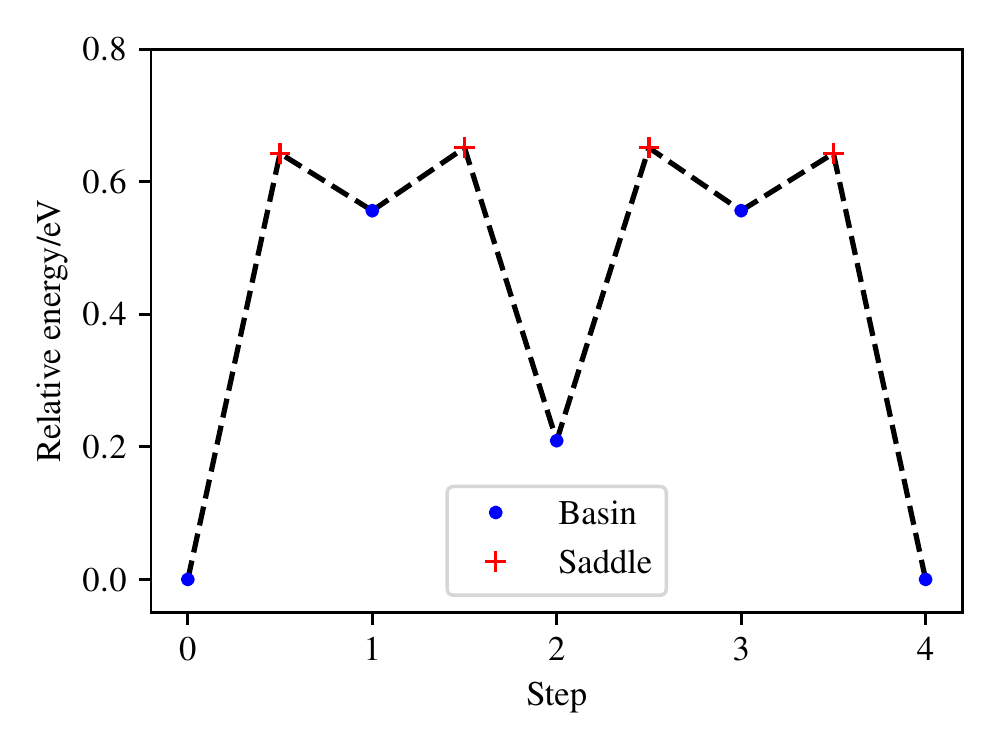}
    } \hfill
    \subfloat[\label{fig:v2_hi}$\text{V}_2$ via $1$\textsuperscript{st}~NN]{
        \includegraphics[width=0.32\textwidth]{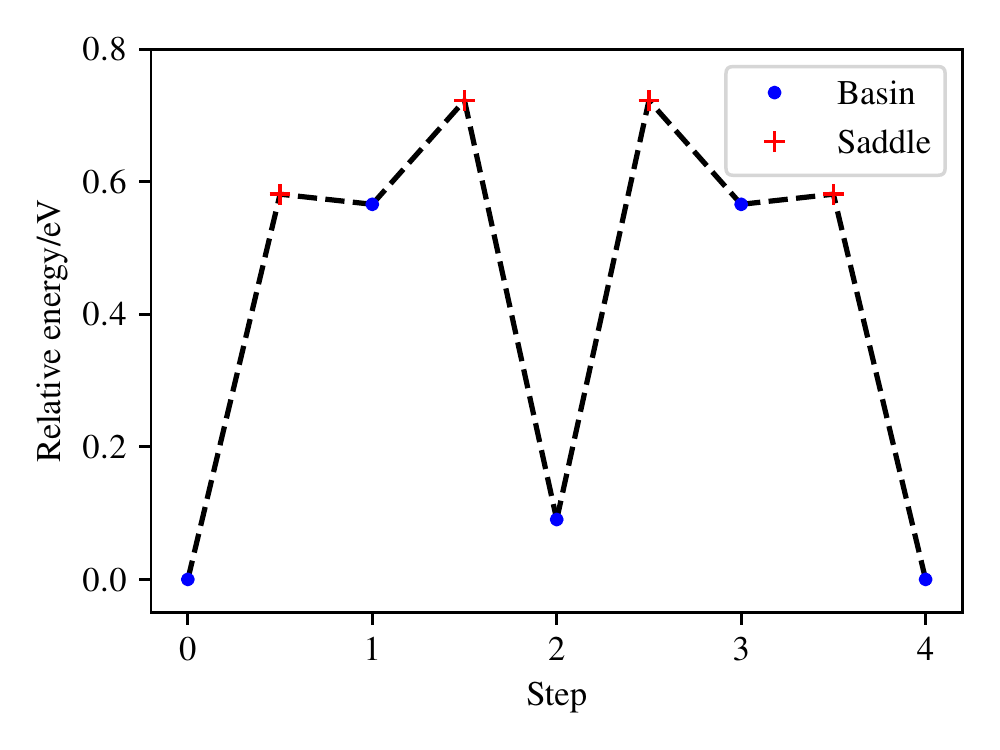}
    }

    \subfloat[\label{fig:v3}$\text{V}_3$]{
        \includegraphics[width=0.32\textwidth]{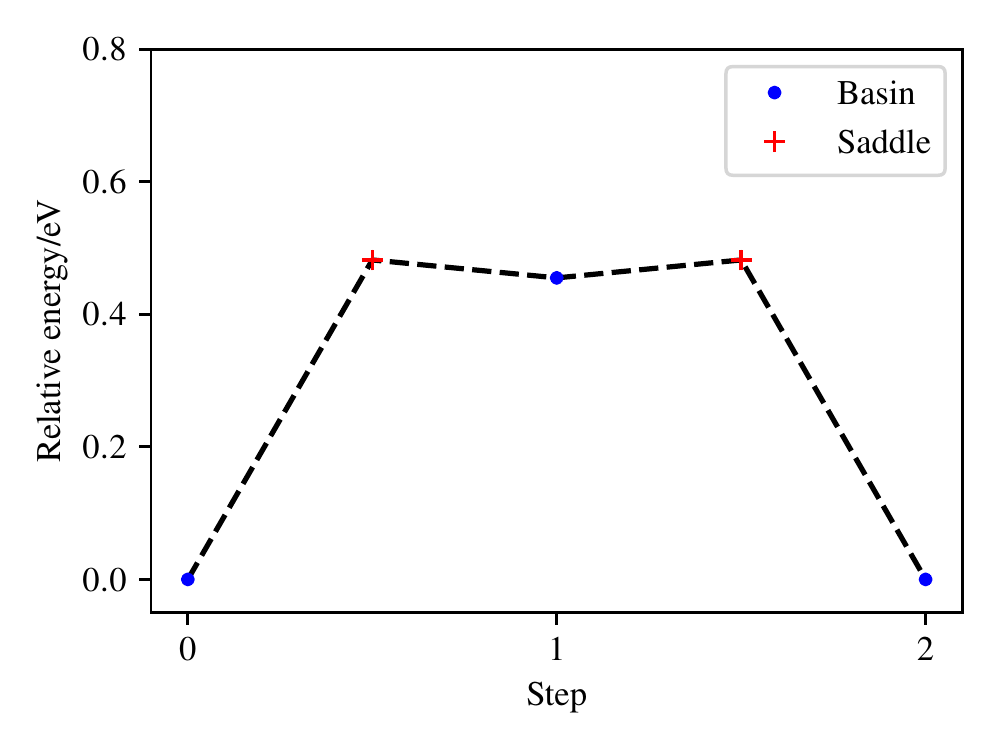}
    } \hfill
    \subfloat[\label{fig:v4}$\text{V}_4$]{
        \includegraphics[width=0.32\textwidth]{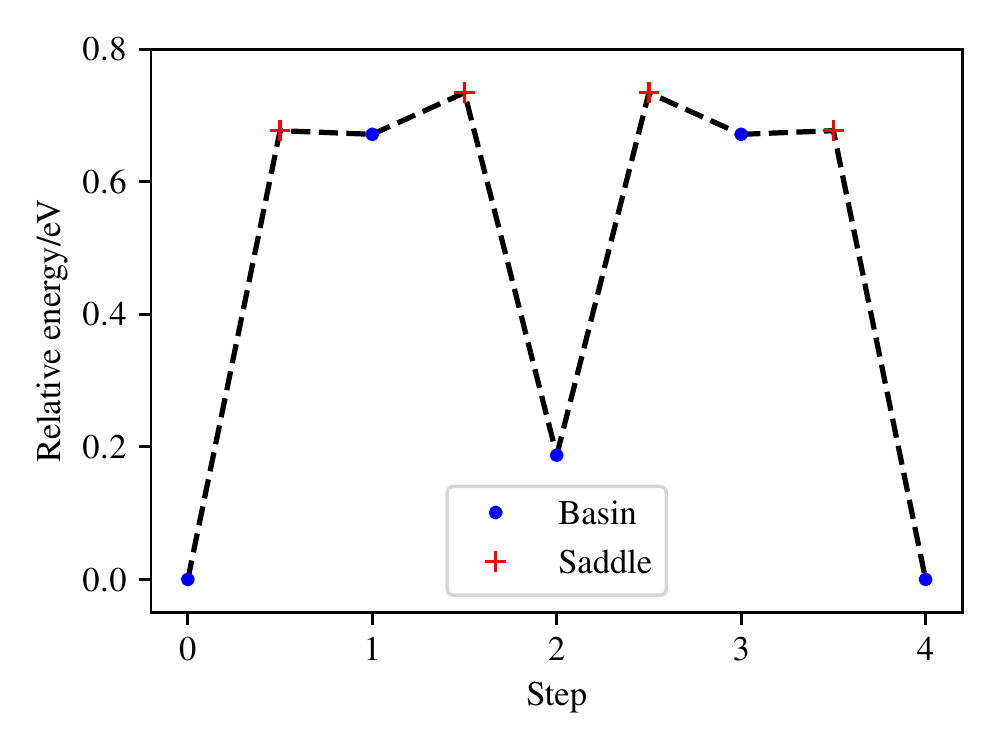}
    } \hfill
    \subfloat[\label{fig:v5}$\text{V}_5$]{
        \includegraphics[width=0.32\textwidth]{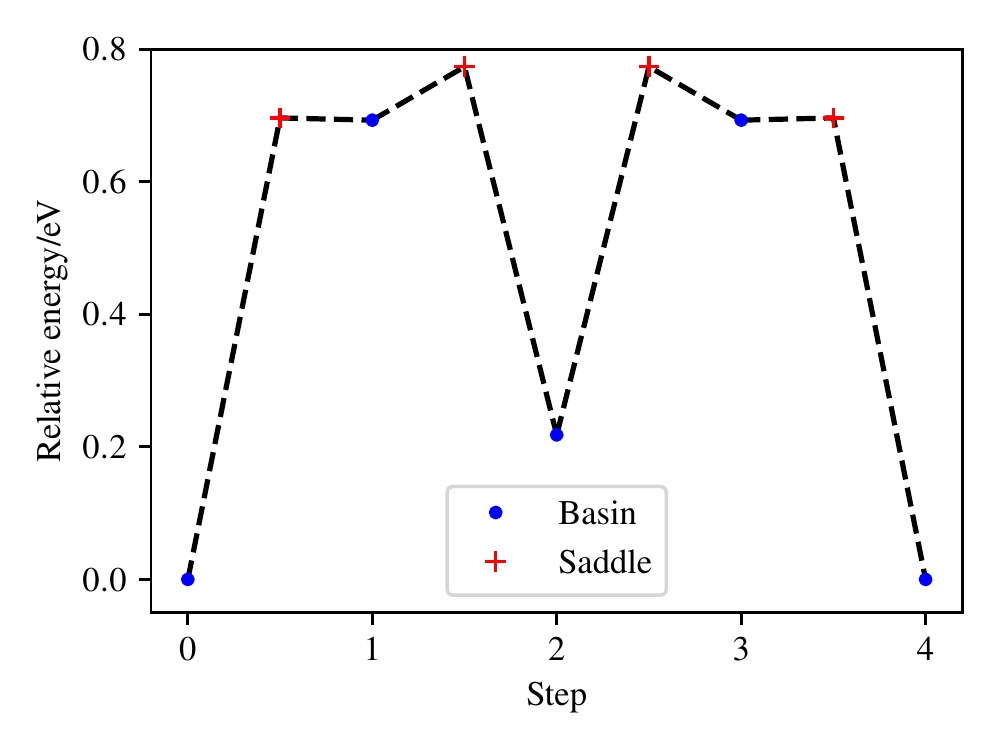}
    }

    \caption{\label{fig:cluster_prof}Energy profiles for the vacancy cluster diffusion mechanisms sketched in \cref{fig:mech_vnh0} -- extracted from OLKMC simulations at \num{300}\si{\kelvin}. \Cref{fig:v1} has energy barriers: \num{0.64} and \num{0.096}\si{\eV} and corresponding kinetic pre-factors: \num{7.44e13} and \num{1.18e13}\si{\Hz}. \Cref{fig:v2_lo} has energy barriers: \num{0.64}, \num{0.095}, \num{0.44} and \num{0.087}\si{\eV} and corresponding kinetic pre-factors: \num{1.04e14}, \num{1.06e13}, \num{7.52e13} and \num{1.11e13}\si{\Hz}. \Cref{fig:v2_hi} has energy barriers: \num{0.58}, \num{0.16}, \num{0.63} and \num{0.015}\si{\eV} and corresponding kinetic pre-factors: \num{2.97e14}, \num{9.33e12}, \num{1.16e14} and \num{5.16e12}\si{\Hz}. \Cref{fig:v3} has energy barriers: \num{0.48} and \num{0.027}\si{\eV} and corresponding kinetic pre-factors: \num{5.22e13} and \num{5.24e12}\si{\Hz}. \Cref{fig:v4} has energy barriers: \num{0.68}, \num{0.063}, \num{0.55} and \num{0.0056}\si{\eV} and corresponding kinetic pre-factors: \num{3.41e13}, \num{7.62e12}, \num{3.94e13} and \num{3.01e12}\si{\Hz}. \Cref{fig:v5} has energy barriers: \num{0.70}, \num{0.081}, \num{0.54}, \num{0.37}, \num{0.56} and \num{0.0034}\si{\eV} and corresponding kinetic pre-factors: \num{3.12e13}, \num{7.17e12}, \num{6.00e13}, \num{6.96e12}, \num{4.31e13} and \num{2.58e12}\si{\Hz}.}
\end{figure*}

\begin{figure*}[t]
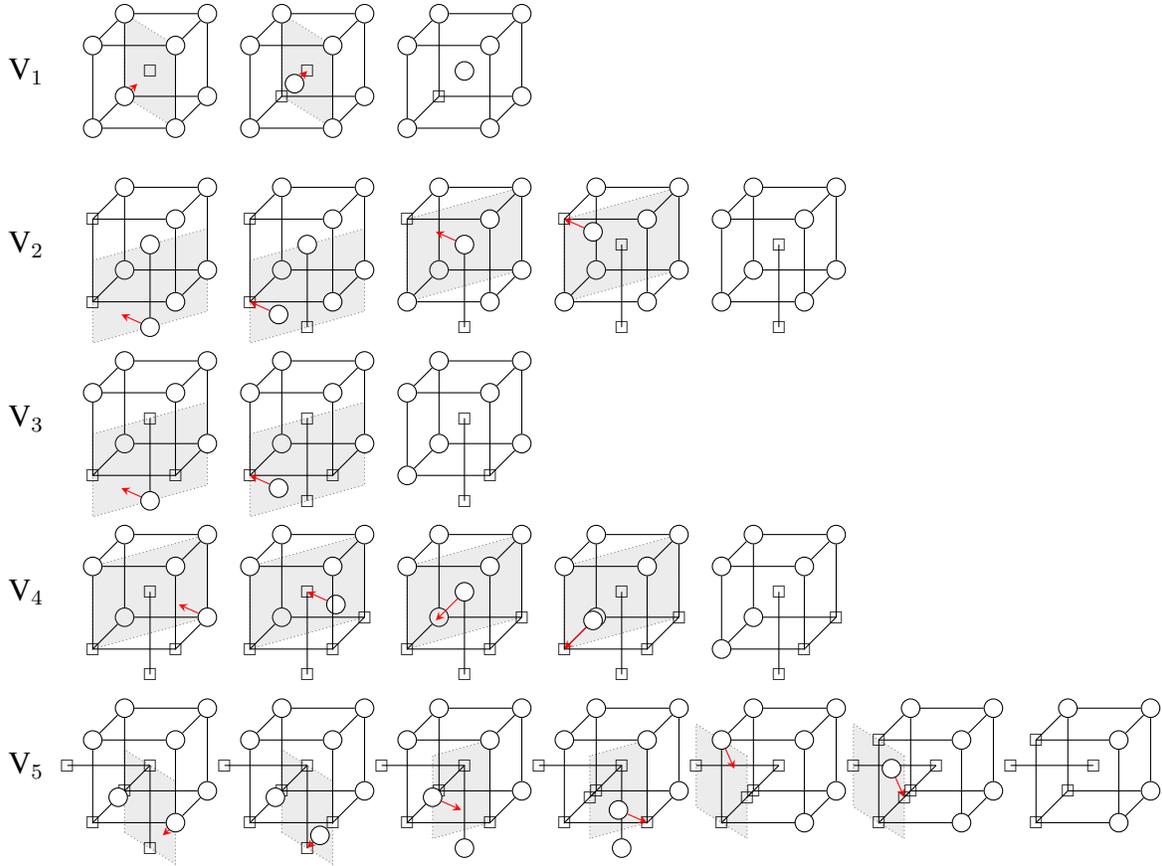

    \centering



    \caption{
        Diffusion mechanisms for vacancy-cluster diffusion in the $\alpha$--Fe lattice. White circles represent an occupied lattice site; small $\square$ symbols indicate an unoccupied BCC lattice site; arrows mark the path of an atom during a mechanism and transparent grey planes act as a guide to the eye containing the atomic path. Small perturbations away from lattice sites have been omitted for clarity. See \cref{fig:cluster_prof} for the corresponding energy profiles. Note for $\text{V}_2$ only the lower energy mechanism (\cref{fig:v2_lo}) has been sketched.}
    \label{fig:mech_vnh0}
\end{figure*}

The $\text{V}_n$ diffusion results are presented in \cref{fig:cluster_diff} and summarised in \cref{tab:cluster_diff}. The energy profiles for the identified mechanisms are presented in \cref{fig:cluster_prof} alongside the mechanisms themselves in \cref{fig:mech_vnh0}. In \cref{fig:cluster_diff}, we see convergence to diffusive behaviour for all clusters, verifying that we are reaching diffusive timescales. These timescales are a property of each system and range between \num{e-5}\si{\second} and \num{1}\si{\second}. The expected behaviour for cluster diffusivity is for larger clusters to become less mobile. This trend is visible in \cref{tab:cluster_diff} but, the diffusivity for $\text{V}_3$ seems to reverse this trend. We shall now discuss each cluster in detail to fully understand this behaviour.

\paragraph*{$\text{V}_1$}The single vacancy diffuses by $\frac{1}{2}\langle 111 \rangle$ vacancy hops (a feature common to all the clusters and complexes) with an activation energy of $0.65$\si{\eV} and kinetic pre-factor \num{7.44e13}\si{\Hz}. The energy barrier and mechanism are in good agreement with the literature~\cite{Messina2015, EcheverriRestrepo2020}.

\paragraph*{$\text{V}_2$}The minimum-energy configuration (MEC) for $\text{V}_2$ is the second nearest neighbour~(NN) pair followed by the \num{1}\textsuperscript{st}~NN then \num{4}\textsuperscript{th}~NN orientations, this matches the literature~\cite{Djurabekova2010}. The predominant $\text{V}_2$ diffusion mechanism was oscillations between the $2$\textsuperscript{nd}~NN and $4$\textsuperscript{th}~NN states, with an energy barrier of \num{0.65}\si{\eV}. The \num{1}\textsuperscript{st}~NN pathway may be expected to be the dominant mechanisms, as one may predict the transition to the lower-energy $1$\textsuperscript{st}~NN state to have a lower energy-barrier. However, the  $2$\textsuperscript{nd}~NN to $1$\textsuperscript{st}~NN transition has an energy barrier of $0.72$\si{\eV}, making it kinetically less-favourable. The $\text{V}_2$ diffusion barrier is very close to the diffusion barrier for $\text{V}_1$, this may be an artefact of the EAM potential used, as \textit{ab initio} studies typically predict an energy barrier $0.05$--$0.11$\si{\eV} lower~\cite{Djurabekova2007,Fu2004}. Nevertheless, this agrees with other works that use similar semi-empirical potentials~\cite{Messina2015} hence, this discrepancy is an artefact of the potential and would be resolved with improved potentials. $\text{V}_2$ has a diffusivity approximately half $\text{V}_1$, this is due to the combination of a near-identical energy barrier but requiring two vacancy-hops to diffuse.

\paragraph*{$\text{V}_3$}As previously hinted, $\text{V}_3$ defies the expectation and is the most mobile cluster with a diffusivity more than an order of magnitude higher than $\text{V}_1$ at $300$\si{\kelvin}. This is due to the MEC permitting a vacancy hop with an energy barrier of $0.48$\si{\eV}, that almost immediately reforms the MEC just displaced/rotated. This means, similarly to $\text{V}_1$, $\text{V}_3$ can diffuse without changing its shape. The simulated MEC matches theoretical predictions, as does the mechanism~\cite{Fu2004}.

\paragraph*{$\text{V}_4$}The mobility of $\text{V}_4$ resumes the decreasing trend. This is predominantly due to the high energy barrier, $0.73$\si{\eV}, required to break apart the MEC. \Refcite{Fu2004} used DFT to compute the $\text{V}_4$ diffusion mechanism, they obtained a lower energy barrier of $0.48$\si{\eV} however, their mechanism matches ours. This could be due to image interactions introduced by the small $4^3$ unit-cell supercell used or indicate work is required on the EAM potential. Nevertheless, the match in mechanism pathway is reassuring and indicates the key physics is being captured by the potential.

\paragraph*{$\text{V}_5$}Similarly to $\text{V}_4$, $\text{V}_5$ continues to become less mobile as its size increases.  This is again predominantly due to the high energy barrier, now $0.77$\si{\eV}, required to break apart the MEC. Furthermore, additional steps are required to diffuse, increasing the probability of backtracking.

\subsubsection{Discussion}

Off-lattice KMC has successfully been applied to study the diffusion of vacancy clusters. The mechanisms predicted and diffusivity-trends match those seen in the literature~\cite{Djurabekova2010, Fu2004, Messina2015}. However, they have all been predicted, without \emph{a priori} assumptions, by the highly general OLKMC framework. This is exemplified by the counter-intuitive diffusion mechanisms of $\text{V}_2$ that could easily be misidentified if using simpler models \eg final-to-initial-state-energy (FISE)~\cite{Messina2015}. Although, this could have been captured with traditional KMC and careful DFT analysis, here it arises naturally without any special consideration or bias. The range in kinetic pre-factors, spanning almost two orders of magnitude, emphasises the need to compute pre-factors even in single element systems.

During our experiments at \num{300}\si{\kelvin}, no dissociation of the vacancy clusters occurred. This confirms the dissociation barrier is higher than the diffusion barrier, which is in line with the literature~\cite{Messina2015}. Exploring diffusion across a range of (higher) temperatures with OLKMC would offer an opportunity to confirm the energy barriers with a fit to the Arrhenius equation and study the dissociation behaviour. If larger clusters become increasingly immobile, dissociation may become the predominant form of diffusion.

\subsection{\label{sec:complex}\texorpdfstring{$\text{V}_n\text{H}$}{Hydrogen vacancy-cluster} complex diffusion}

To build upon the cluster diffusivity results, we add a single H atom into each of the clusters described in \cref{sec:cluster}, forming $\text{V}_n\text{H}$ complexes. The cluster acts as a trap for the H atom, which can then detrap, diffuse rapidly through the lattice and re-trap at another (possibly the same) cluster. Alternatively, the complex itself can diffuse. We refer to the H trapping sites within the vacancy clusters as \emph{deep} trapping-sites.

\begin{figure}[tb]
    \centering
    \includegraphics[width=\dyncolwidth]{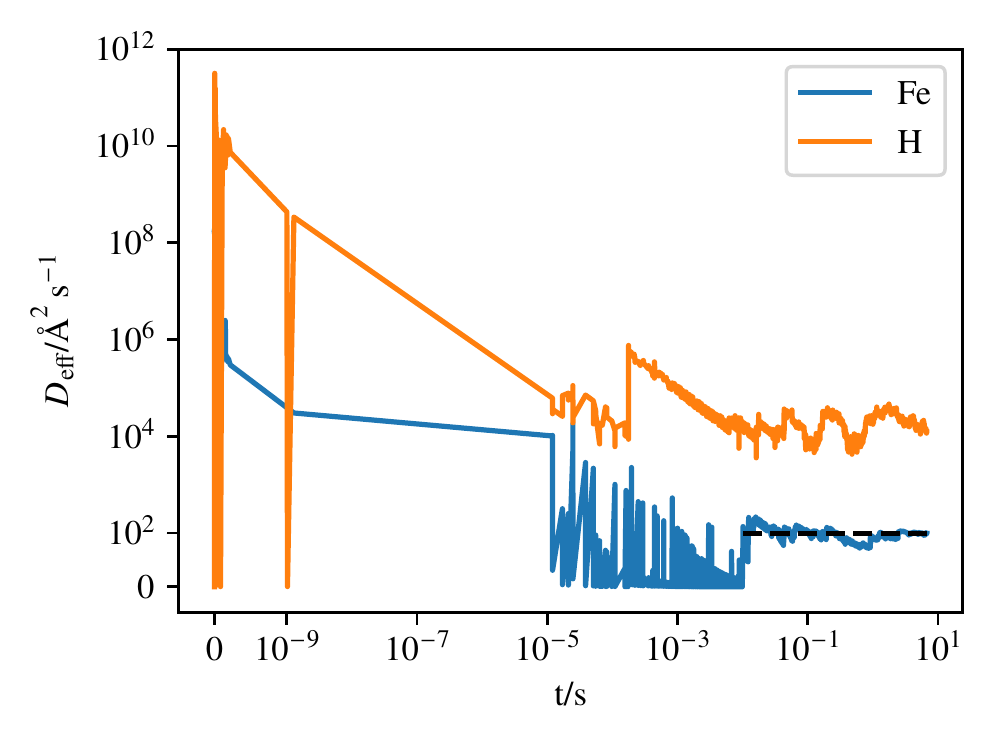}
    \caption{\label{fig:complex_diff}$\text{V}_2\text{H}$ complex diffusing in a perfect $6^3$ unit-cell supercell at $300$\si{\kelvin}, dashed line is a fit to a constant.}
\end{figure}

The parallel of \cref{fig:cluster_diff} is presented in \cref{fig:complex_diff}, for the representative $\text{V}_2\text{H}$ complex. We see timescales range from $10^{-13}$\si{\second} to $10$\si{\second}, this is required to resolve the distinct regimes visible in \cref{fig:complex_diff}. Firstly, below $10^{-9}$\si{\second}, the H atom explores the deep-trapping states within vacancy clusters.  Analytical superbasin acceleration kicks in once all the states are explored, solves the flickering problem between the deep trapping-states and triggers the discontinuity to $10^{-5}$\si{\second} as the H atom escapes then rebinds to the cluster. Next, near $10^{-4}$\si{\second}, the H atom detraps and diffuses to another cluster. Finally, near \num{1e-2}\si{\second}, the complex displaces and beyond converges to diffuse behaviour. Like the experiments in \cref{sec:cluster}, SP searches are only required during the learning phase of the simulations.

\begin{table}[b]
    \centering

    \caption{\label{tab:complex_diff}Summary of cluster-H diffusion results in the $\alpha$--Fe lattice at \num{300}\si{\kelvin}. All diffusivities have a fractional error less than one part in one hundred. Quoted kinetic pre-factors for multi-step mechanisms is that of the highest barrier step.}

    \renewcommand{\arraystretch}{1.3}

    \begin{tabular*}{\columnwidth}{@{}l@{\extracolsep{\fill}}lll@{}}
        \toprule
        Complex              & $\Delta E$/\si{\eV} & $v$/$10^{13}$\si{\Hz} & $D_\text{eff}$/\si{\meter\squared\per\second} \\
        \midrule
        $\text{V}_1\text{H}$ & $0.75(8)$           & \num{23.3}            & \num{3.72e-19}                                \\
        $\text{V}_2\text{H}$ & $0.70(5)$           & \num{8.70}            & \num{1.00e-18}                                \\
        $\text{V}_3\text{H}$ & $0.49(1)$           & \num{9.22}            & \num{7.50e-16}                                \\
        $\text{V}_4\text{H}$ & $0.71(2)$           & \num{7.86}            & \num{3.95e-18}                                \\
        $\text{V}_5\text{H}$ & $0.74(8)$           & \num{7.33}            & \num{5.16e-20} \\
        \bottomrule
    \end{tabular*}
\end{table}

\begin{figure*}[t]
    \centering

    \subfloat[\label{fig:v1h1}$\text{V}_1 \text{H}$]{
        \includegraphics[width=0.32\textwidth]{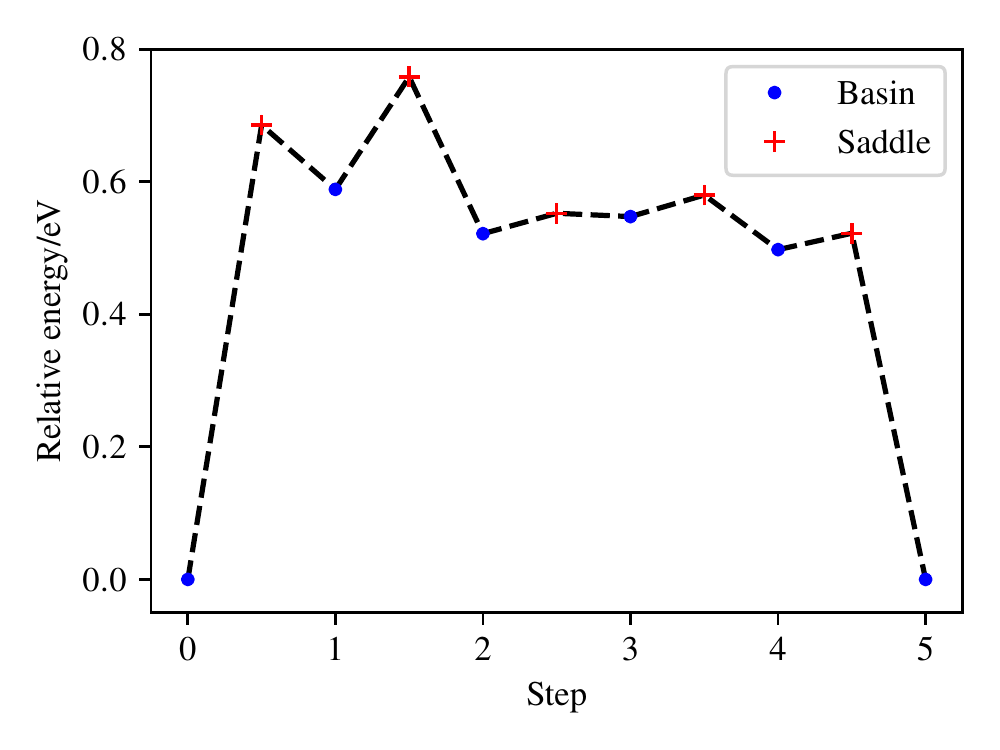}
    } \hfill
    \subfloat[\label{fig:v2h1}$\text{V}_2 \text{H}$]{
        \includegraphics[width=0.32\textwidth]{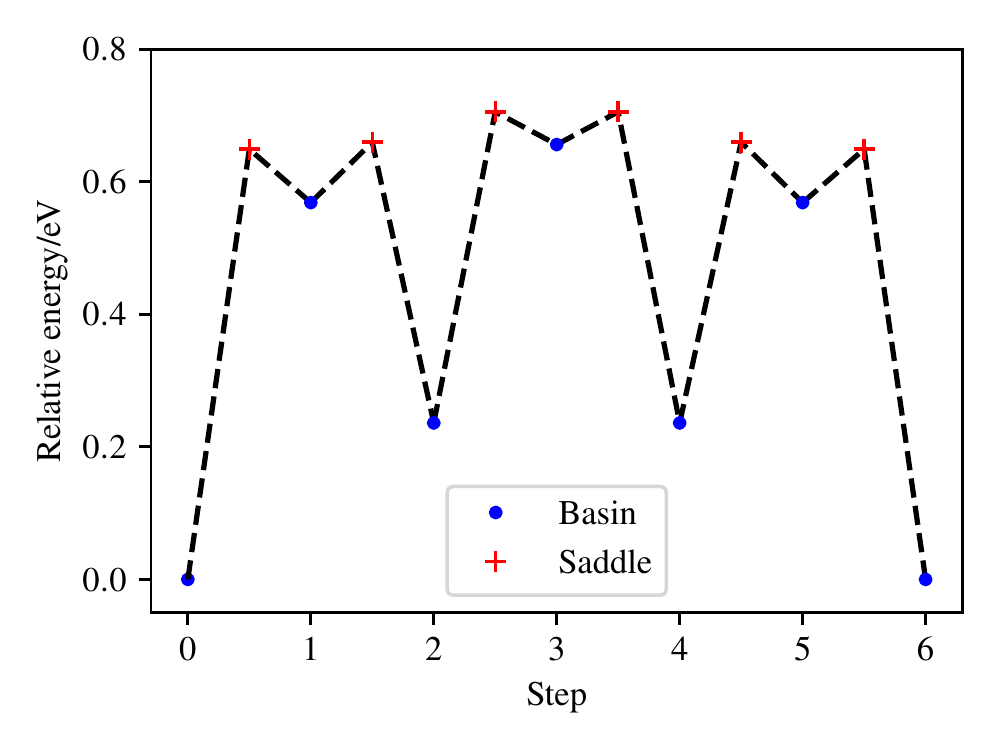}
    } \hfill
    \subfloat[\label{fig:v3h1}$\text{V}_3 \text{H}$]{
        \includegraphics[width=0.32\textwidth]{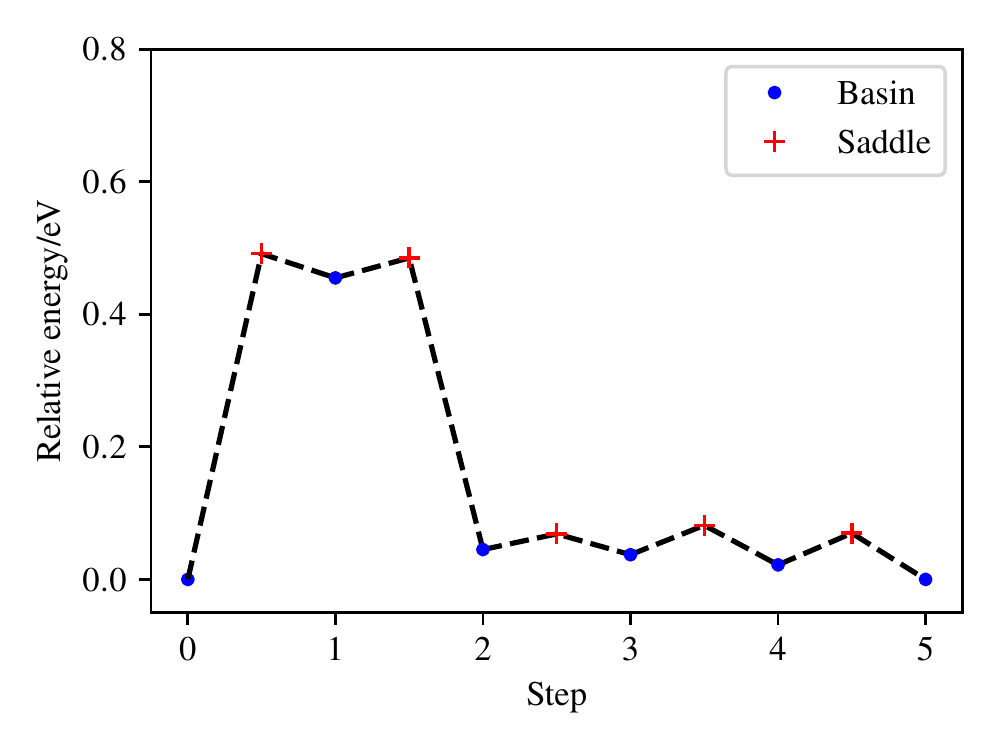}
    }

    \subfloat[\label{fig:v4h1}$\text{V}_4 \text{H}$]{
        \includegraphics[width=0.48\textwidth]{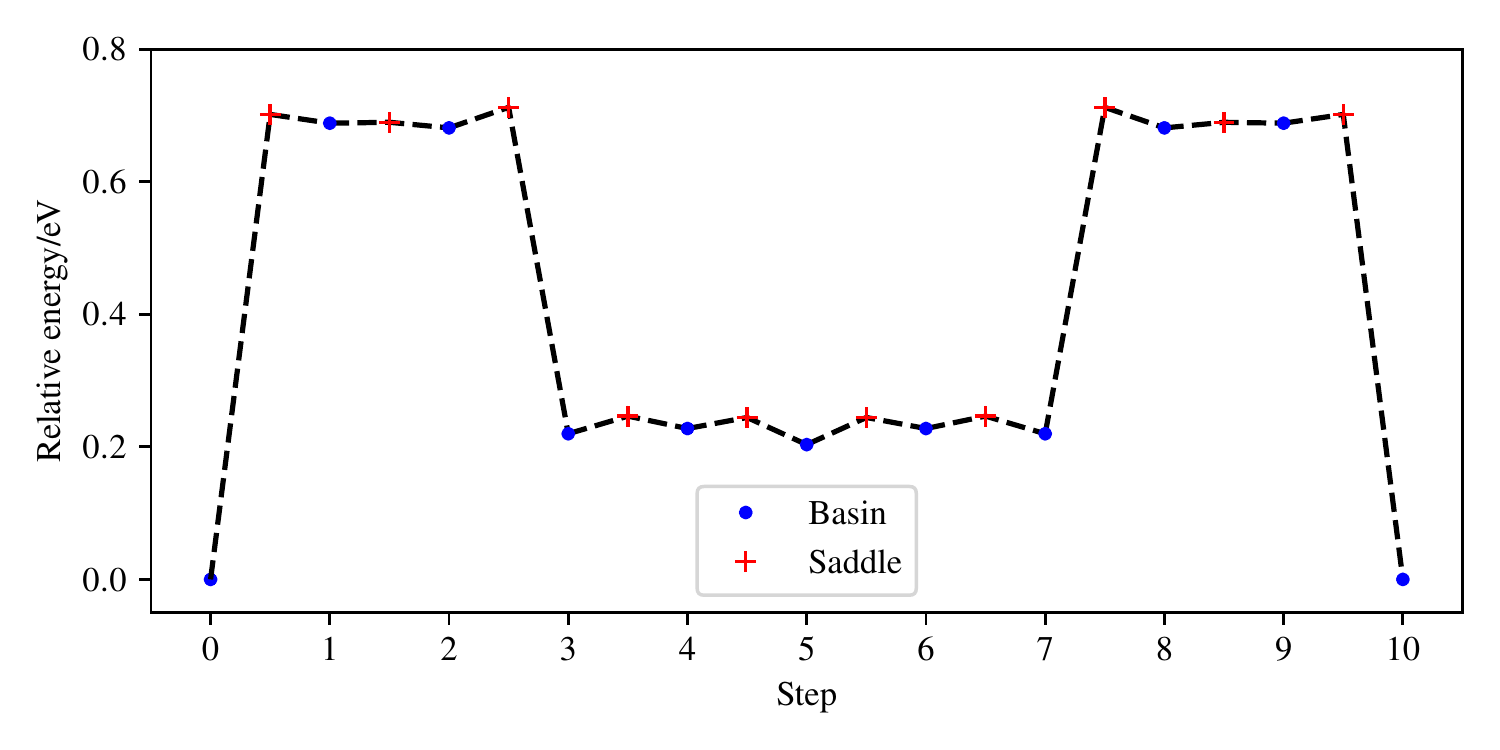}
    } \hfill
    \subfloat[\label{fig:v5h1}$\text{V}_5 \text{H}$]{
        \includegraphics[width=0.48\textwidth]{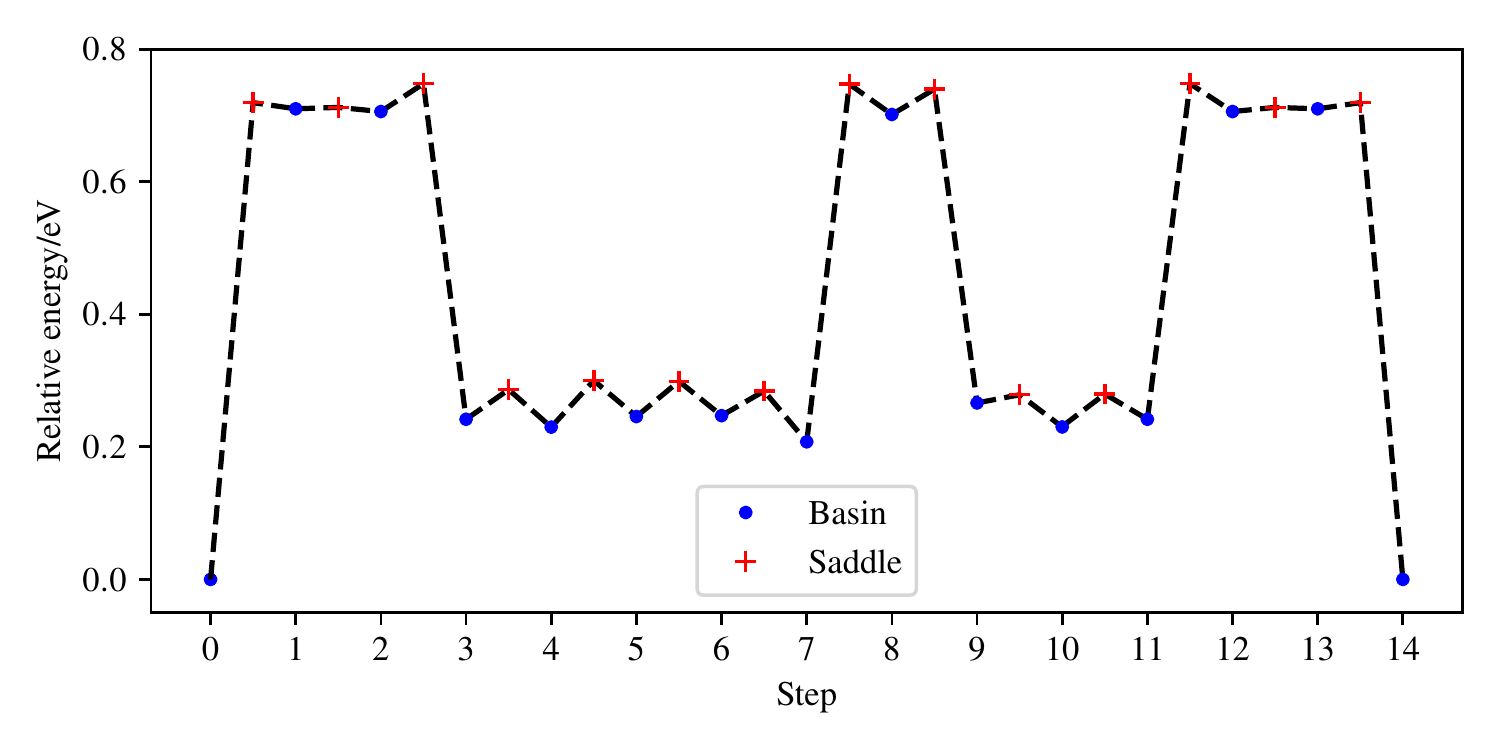}
    }
    \caption{\label{fig:complex_prof}Energy profiles for the cluster-H diffusion mechanisms sketched in \cref{fig:mech_vnh1} -- extracted from OLKMC simulations at \num{300}\si{\kelvin}. The individual barriers and kinetic pre-factors are omitted for brevity. }
\end{figure*}

\begin{figure*}
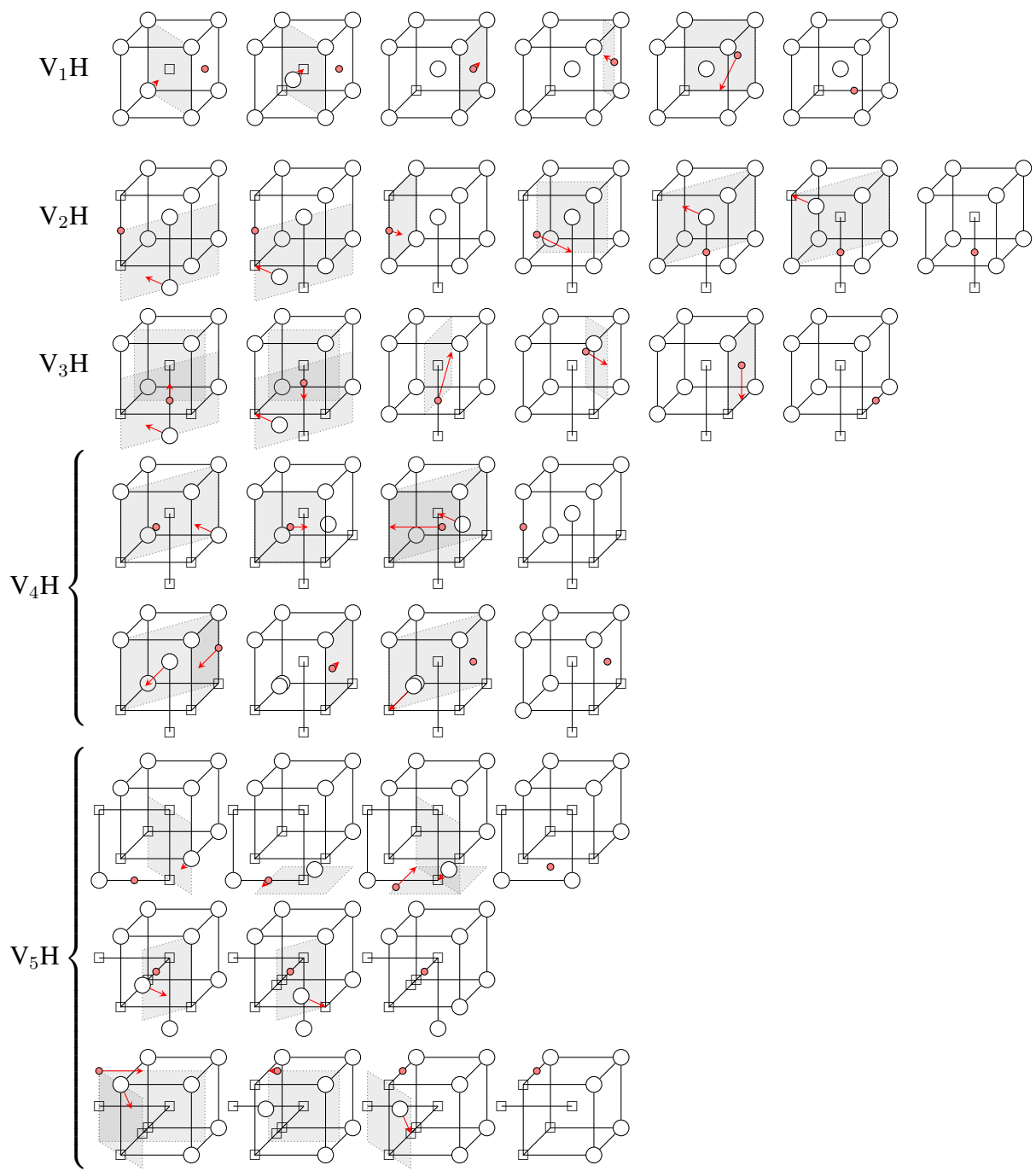

    \centering



    \caption{Diffusion mechanisms for $\text{V}_n\text{H}_1$ complexes in the $\alpha$--Fe lattice. Small red circles mark a H atom; white circles represent an occupied lattice site; small $\square$ symbols indicate an unoccupied BCC lattice site; arrows mark the path of an atom during a mechanism and transparent grey planes act as a guide to the eye containing the atomic path. Small perturbations away from lattice/octahedral/tetrahedral sites have been omitted for clarity. See \cref{fig:complex_prof} for the corresponding energy profiles. Note for $\text{V}_4\text{H}$ only the frames corresponding to steps \num{0}--\num{3} and \num{7}--\num{10} in \cref{fig:v4h1} are shown and similarly for $\text{V}_5\text{H}$ with steps \num{0}--\num{2}, \num{7}--\num{9} and \num{18}--\num{20} from \cref{fig:v5h1}. }
    \label{fig:mech_vnh1}
\end{figure*}

The complex's diffusivities and energy barriers are summarised in \cref{tab:complex_diff}; the energy profiles for the identified mechanisms are presented in \cref{fig:complex_prof} and the mechanisms themselves in \cref{fig:mech_vnh1}. We expect higher energy barriers for small complexes, especially $\text{V}_1\text{H}$, due to the larger relative steric hindrance provided by the H atom in the smaller clusters. As the clusters get larger, one would predict that the `regular' H-free mechanisms could occur whilst the H atom occupies the opposite side of the cluster, interacting minimally. Hence, larger complexes are expected to diffuse similarly to their corresponding clusters. Interestingly, we see for the larger clusters investigated that the introduction of H lowers the energy barrier, indicating a different mechanism may be occurring. We shall now discuss each complex in detail to fully understand this behaviour.

\begin{figure}[tb]
    \centering
    \includegraphics[width=\dyncolwidth]{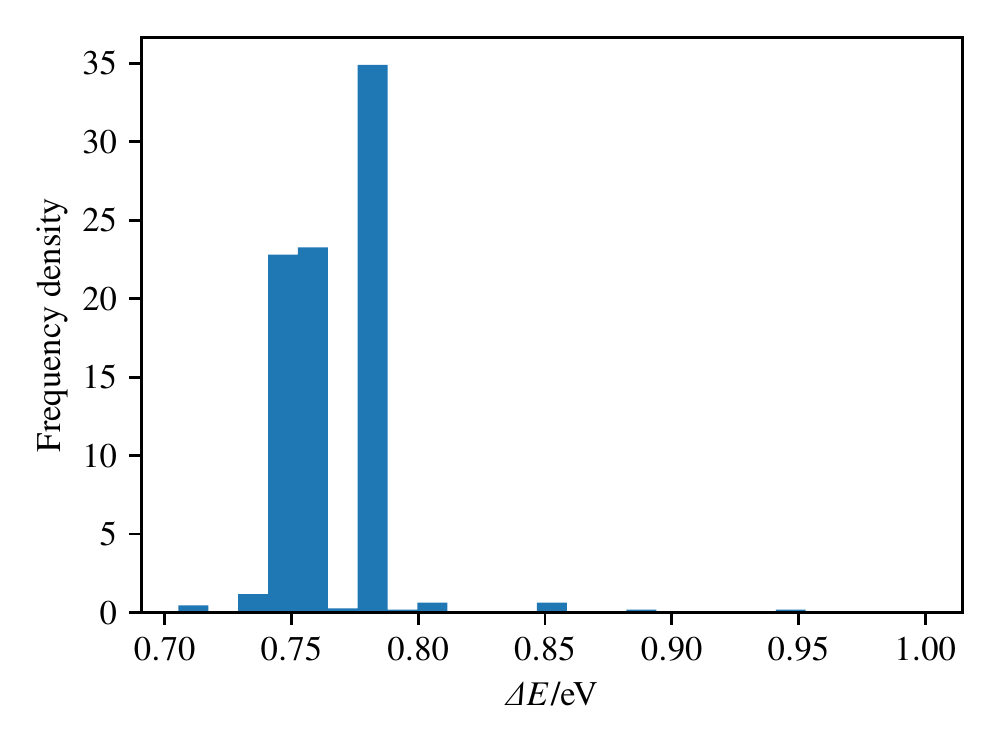}
    \caption{\label{fig:hist}Histogram of energy barriers for $\text{V}_2\text{H}$ complex diffusion in a perfect $6^3$ unit-cell supercell at $300$\si{\kelvin}.}
\end{figure}

\paragraph*{$\text{V}_1\text{H}$}As predicted by the steric argument, $\text{V}_1\text{H}$ experiences the largest increase in diffusion barrier. The H atom can occupy \num{6} deep trapping-sites in the vacancy, close to the adjacent octahedral sites. The H atom must be `pushed' out of the vacancy by an Fe atom during a $\frac{1}{2}\langle 111 \rangle$ hop which increases the energy barrier, demonstrating co-dependence. The partially-escaped H atom then rebinds with the newly formed vacancy. This mechanism is asymmetric; from the perspective of the reverse direction, the H atom first escapes and then pushes a Fe atom into the vacancy. The energy barrier, energy profile and mechanism closely match the EAM results of \refcite{EcheverriRestrepo2020}, validating that our norm-based classification generalises to small interstitials and heterogeneous systems. In the literature it is suggested that the reverse direction is the preferred mechanism~\cite{Hayward2013, EcheverriRestrepo2020} however, in our experiments this only occurs $\approx 30$\% of the time. This could be due to a kinetic bias towards the forward direction at \num{300}\si{\kelvin}.

\paragraph*{$\text{V}_2\text{H}$}The H atom can occupy \num{14} deep trapping-sites in the $\text{V}_2$ \num{2}\textsuperscript{nd}~NN cluster, again close to the surrounding octahedral sites. The larger $\text{V}_2\text{H}$ complex has a diffusion mechanism more similar to the H free case than $\text{V}_1\text{H}$; the vacancies first move into a $4$\textsuperscript{th}~NN coordination leaving the H atom in the original vacancy, the H atom then jumps out and into the moved vacancy. Finally, the original vacancy hops and reforms a displaced $2$\textsuperscript{nd}~NN complex. Less interaction between the H atom and moving Fe atom(s) results in a lower total energy barrier of $0.71$\si{\eV}, compared to $\text{V}_1\text{H}$. This explains the corresponding ten-fold increase in diffusivity compared to $\text{V}_1\text{H}$. While for $\text{V}_1\text{H}$ only one diffusive mechanism occurred, for $\text{V}_2\text{H}$ a group of diffusive mechanisms, with energy barriers \num{0.71}--\num{0.95}\si{\eV} were found. \Cref{fig:hist} depicts the distribution of diffusion-mechanism energy-barriers identified. The most prolific mechanism at $300$\si{\kelvin} had an energy barrier of \num{0.78}\si{\eV}. The minimum barrier mechanism was probably suppressed by the high likelihood of the vacancy hopping back to the initial configuration before the H atom could hop to the new vacancy. In contrast the \num{0.78}\si{\eV} mechanism progressed through a \num{1}\textsuperscript{st}~NN intermediate, which is equally likely to move to a new state or backtrack as the H atom can freely move between the two vacancies. This highlights the significance of kinetics vs energetics at room temperature.

\paragraph*{$\text{V}_3\text{H}$}The introduction of H to $\text{V}_3$ has little effect on the diffusion mechanisms, only fractionally increasing the exceptionally low-barrier $\text{V}_3$ mechanism to $0.49$\si{\eV}. This results in $\text{V}_3\text{H}$ not disassociating for the entirety of the simulation. The $\text{V}_3$ cluster provides $18$ deep trapping-sites for H.  Similarly to $\text{V}_2\text{H}$, a range of diffusion mechanisms were identified with energy barriers \num{0.49}--\num{0.57}\si{\eV}. This time the minimum energy mechanism accounted for a substantial fraction of the observed mechanisms but, was still not the most frequent.

\paragraph*{$\text{V}_4\text{H}$}The $\text{V}_4$ cluster provides $20$--$32$ deep trapping-sites for H (depending on the cluster configuration). For the first time, introduction of an H atom decreases the energy barrier and increases the diffusivity of the complex, compared to the cluster alone. Studying \cref{fig:mech_vnh0} and \cref{fig:mech_vnh1}, we see the motion of Fe atoms remains the same as $\text{V}_4$. Although the H atom has extra space in the large cluster to `avoid' the hopping Fe atom, as predicted by the steric argument, instead it remains close and lowers the energy of the saddle-point(s). This effect can be likened to the pathway provided by the partially-escaped H atom pushing a Fe atom in $\text{V}_1\text{H}$. However, due to the increased size of the cluster and its shape, the H atom does not need to escape before providing this push. Furthermore, the intermediate Fe configuration (steps \num{3}--{7} in \cref{fig:v4h1}) is connected, meaning the H atom can easily reach all the deep traps in the cluster (unlike $\text{V}_2\text{H}$) hence, the forwards and reverse directions are equally likely.

\paragraph*{$\text{V}_5\text{H}$}The $\text{V}_5$ cluster provides $28$--$36$ deep trapping-sites for H and continues the $\text{V}_4\text{H}$ trend of decreasing the energy barrier compared to its corresponding cluster. In \cref{fig:mech_vnh1} we see (again contrary to the steric hypothesis) the H atom remains close to the hopping Fe atom, with the first and last hops very similar to the corresponding steps of the $\text{V}_4\text{H}$ mechanism. Despite the lower energy barrier, $\text{V}_5\text{H}$ diffusivity is reduced compared to $\text{V}_5$. This can be attributed to the additional complexity of the mechanism (visible in \cref{fig:v5h1}) and additional backtracking, all of which reinforces the importance of kinetic effects at room temperature. To the authors knowledge, this is the first-time mechanisms for $\text{V}_5\text{H}$ have been reported.

\subsubsection{Discussion}

\begin{figure}[tb]
    \centering
    \includegraphics[width=\dyncolwidth]{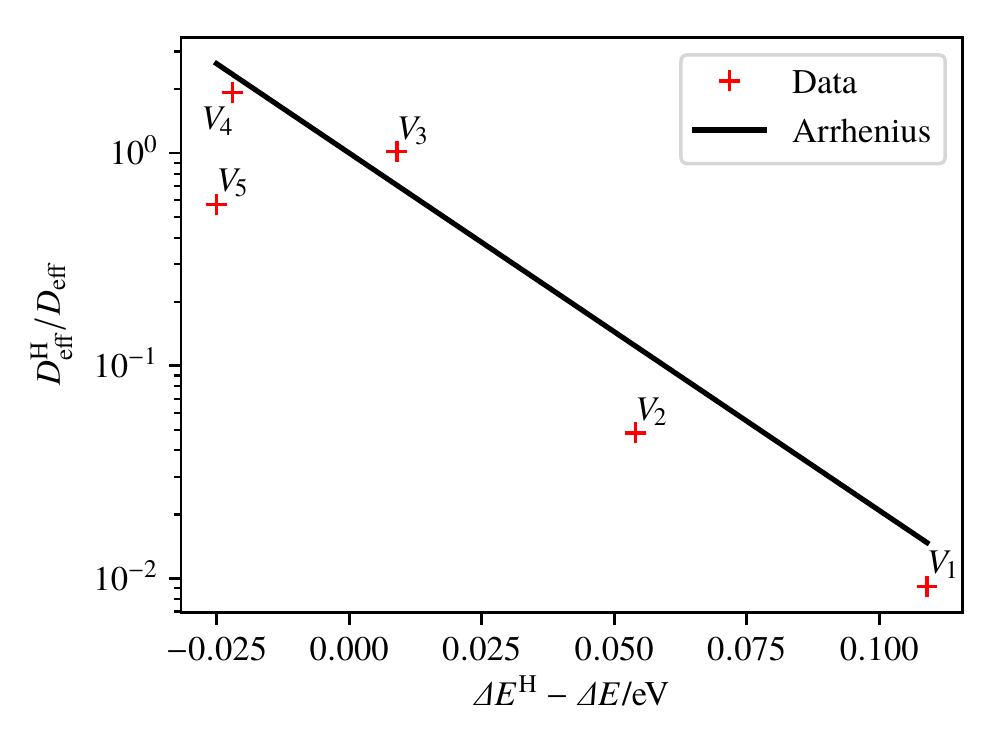}
    \caption{\label{fig:arr_expect}Comparison between $\text{V}_n$ and $\text{V}_n\text{H}$ diffusion barriers vs their corresponding effective diffusivities -- expectation is \cref{equ:expect} where we assume equal Arrhenius pre-factors.}
\end{figure}

As we move from $\text{V}_n$ clusters to $\text{V}_n\text{H}$ complexes, we may expect the diffusivities to scale with the energy barriers according to the classical Arrhenius behaviour:
\begin{align}
    \frac{D^\text{H}_\text{eff}}{D_\text{eff}} \simeq e ^ {-\beta \left( \Delta E^\text{H} - \Delta E \right)} \label{equ:expect}
\end{align}
with $\beta = \frac{1}{k_B T}$ and superscripts denoting the complexes. This assumes both the cluster and the complex have the same Arrhenius pre-factor, this assumption can be motivated by the equivalent pathways of Fe atoms during diffusion in \cref{fig:mech_vnh0} and \cref{fig:mech_vnh1}. The comparison is drawn in \cref{fig:arr_expect}, where we see a general conformance to expectation. We see the largest deviations for $\text{V}_2\text{H}$ and $\text{V}_5\text{H}$, these complexes require three high-barrier steps during their diffusion mechanism hence, these deviations are probably due to the increased likelihood of backtracking.

For all the complexes beyond $\text{V}_1\text{H}$, a number of alternative mechanisms were accessible at \num{300}\si{\kelvin}. Off-lattice KMC was able to discover these on-the-fly; traditional KMC simulation that use (typically small) pre-determined lists of mechanisms could easily omit mechanisms that contribute to interesting behaviour. Capturing only the lowest energy barrier mechanisms may not be sufficient. \Cref{fig:hist} exemplifies this for $\text{V}_2\text{H}$, which is the simplest example of this increasing complexity. This effect will become worse at higher temperatures where alternate, higher-energy mechanisms become ever more common.

Similarly to \cref{sec:cluster}, the vacancies in the complexes did not dissociate during the simulation. This suggests the dissociation barrier remains higher than the complex's diffusion barriers but, sheds little light on the effect of H on the dissociation barrier. Extending the simulations across a range of temperatures would offer an opportunity to confirm the energy barriers with a fit to the Arrhenius equation and (through comparisons with the H free case) study the effect of H on the dissociation barrier.

It has also been found that the interaction between H and vacancy-clusters has not followed the pattern if extrapolating from $\text{V}_1\text{H}$; perhaps the most interesting behaviour is the reduction in diffusion barrier, upon introduction of H into the larger clusters $\text{V}_4\text{H}$ and $\text{V}_5\text{H}$. A similar phenomenon, dubbed the \emph{hydrogen lubrication effect}, has been explored computationally in FCC metals~\cite{Du2020}. If this trend continues and the gap continues to increase, it could help to explain the experimentally observed H-induced nano-void migration~\cite{Arakawa2021} and could have implications for HE by contributing towards vacancy agglomeration during the HESIV mechanism~\cite{Nagumo2004, Li2015, Nagumo2019}. A similar atomic mechanism that lowers the diffusion barrier for Vacancy-H complexes could be responsible for the predicted increase in dislocation velocity/mobility~\cite{Gong2020, Katzarov2017}, which could directly support the HELP mechanism~\cite{Birnbaum1994} of HE. As ever, we should be wary of extrapolating; nano-sized cluster and dislocations could be simulated with OLKMC and would need to be studied before drawing conclusions about macroscopic phenomena.

\subsection{Effective hydrogen diffusivity vs classical approaches}

Alongside the diffusion of the complexes in \cref{sec:complex}, we also investigate the motion of the H atom between the network of defects. To explore these results we make comparisons with \citeauthor{Oriani1970}'s theory of equilibrium-trapping~\cite{Oriani1970}, which predicts the effective diffusivity of H, $D_\text{Or}$, is related to the diffusivity in the perfect lattice, $D$, via~\cite{Bombac2017}:
\begin{align}
    D_\text{Or} = D \frac{n_L \theta_L}{n_L \theta_L + n_x \theta_x (1 - \theta_x)} \label{equ:oriani_diff}
\end{align}
where $\theta_x$, $\theta_L$ are the fractional occupancy of available point-trap and regular lattice sites respectively, and $n_x$, $n_L$ the corresponding number of sites. \Citeauthor{Oriani1970}'s theory assumes the traps are stationary hence, $D_\text{Or}$ only takes account for H displacement while the H atom moves between traps.

To obtain the average of $\theta_x$ and $\theta_L$ over the course of the entire simulation we identify the locations of the deep trapping-sites in each frame and determine their centroid. Then, defining $r$ as the distance from the H atom to the centroid, we partition the frames into two sets: trapped frames with the $r < r_x$, a \emph{cut off} trapping radius, and lattice frames with $r \ge r_x$. We now discuss how $r_x$ can be chosen and give us information about the size of the defect. Associating trapping sites to the location of the H atom at trapped frames and similarly for lattice sites/frames, the total time spent in each partition, $t_x$ and $t_L$, are related directly to $\theta$ and $n$ via:
\begin{align}
    n_x\theta_x  = \frac{t_x}{t_x + t_L}
    \quad \text{and} \quad
    n_L\theta_L = \frac{t_L}{t_x + t_L}
\end{align}
and we assume $n_x$ and $n_L$ are approximately constant. As we expect the H atom to spend most of its time bound to the vacancy cluster, we apply $t_L \ll  t_x$ to obtain:
\begin{align}
    D_\text{Or} =  \frac{t_L}{t_x} \frac{D}{1-\frac{1}{n_x}} \label{equ:oriani_diff_simple}
\end{align}

\begin{figure}[tb]
    \centering
    \includegraphics[width=\dyncolwidth]{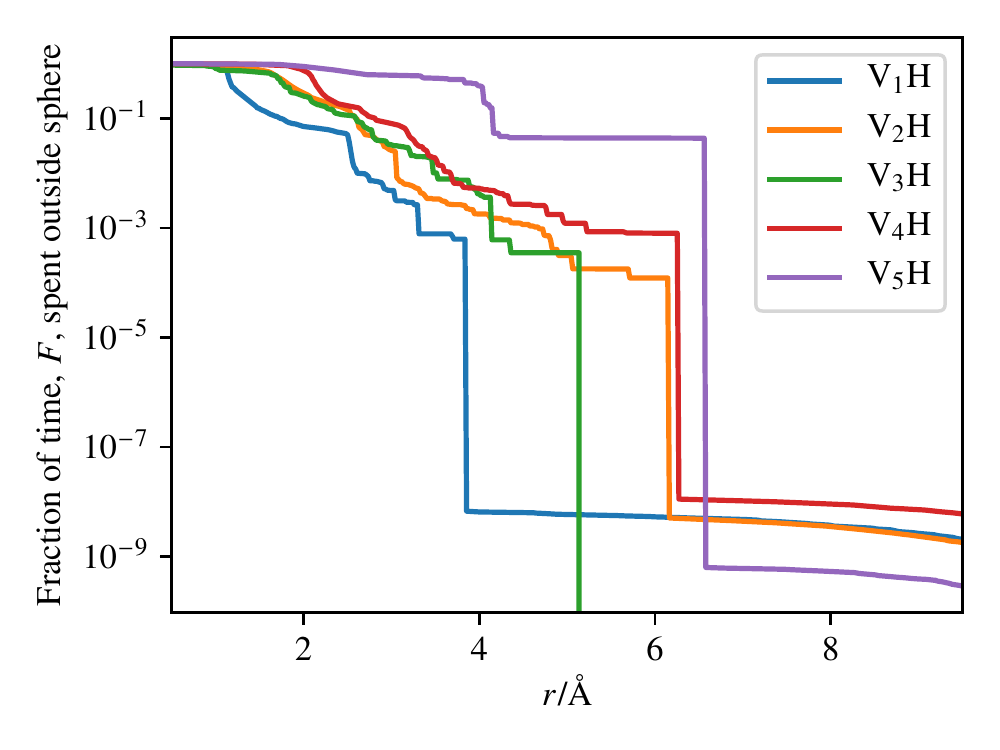}
    \caption{\label{fig:or_plt}Fraction of the total-time the H atom spent outside a sphere of radius $r$ centred on the vacancy cluster at \num{300}\si{\kelvin}.}
\end{figure}

\begin{table*}[b]
    \centering

    \caption{\label{tab:h_energy}Summary of $\text{V}_n\text{H}$ binding energies and detrapping barriers in the $\alpha$--Fe lattice extracted from OLKMC simulations at \num{300}\si{\kelvin}. Additionally, the diffusivity of the H atom is included alongside the trapping radii and \citeauthor{Oriani1970} diffusivities. Note: $\text{V}_3\text{H}$ did not detrap during the simulation hence, the corresponding values are omitted.}

    \renewcommand{\arraystretch}{1.3}

    \begin{tabular*}{\textwidth}{@{}l@{\extracolsep{\fill}}llllll@{}}
        \toprule
        Complex              & $r_x$/\si{\angstrom} & $\Delta E_\text{detrap}$/\si{eV} & $\Delta E - \Delta E_\text{detrap}$/\si{eV} & $E_B$/\si{eV}     & $D_\text{H}$/\si{\meter\squared\per\second} & $D_\text{Or}$/\si{\meter\squared\per\second} \\
        \midrule
        $\text{V}_1\text{H}$ & 3.9                  & \num{0.63(3)}                    & 0.13                                        & \num{0.59}        & $1.57 \pm 0.21 \times 10^{-16}$             & $8.47 \times 10^{-17}$                       \\
        $\text{V}_2\text{H}$ & 6.2                  & \num{0.66(7)}                    & 0.04                                        & \num{0.62}        & $1.25 \pm 0.06 \times 10^{-16}$             & $5.69 \times 10^{-17}$                       \\
        $\text{V}_3\text{H}$ & 5.2                  & -                & -                           & - & $1.24 \pm 0.11 \times 10^{-15}$             & -                            \\
        $\text{V}_4\text{H}$ & 6.3                  & \num{0.70(3)}                    & 0.01                                        & \num{0.65}        & $4.88 \pm 0.80 \times 10^{-16}$             & $1.22 \times 10^{-16}$                       \\
        $\text{V}_5\text{H}$ & 6.6                  & \num{0.70(8)}                    & 0.04                                        & \num{0.66}        & $4.01 \pm 0.27 \times 10^{-17}$             & $6.84 \times 10^{-18}$                       \\
        \bottomrule
    \end{tabular*}
\end{table*}

In order to determine the appropriate value of $r_x$ for each complex, we plot the fraction of the time, $F$, the H atom spent outside a sphere of radius $r$ centred on the vacancy cluster, as a function of $r$. The results are presented in \cref{fig:or_plt}: for small $r$, as we expect, $F\to 1$ as no trap/lattice sites are inside the sphere. Conversely, at large $r$, $F$ levels-out as all the states outside the sphere are linked by approximately-equal low-barrier mechanisms and the H atom must diffuse an approximately constant (but slowly decreasing hence the trail off in \cref{fig:or_plt}) distance before rebinding. Between the two regions is a sharp discontinuity which marks the bounding sphere that contains all the trapping sites hence, this discontinuity defines $r_x$, an effective size for each defect or \emph{trapping atmosphere}. This is recorded in \cref{tab:h_energy} alongside the diffusivity of the H atom and the relevant energy barriers.

\citeauthor{Oriani1970}'s theory was postulated assuming traps as single points in the lattice interacting with H. In the present work, we are able not only to predict $D_\text{eff}$ but also calculate the effective trapping-distance or \emph{atmosphere} of traps with arbitrary structure \ie non-point traps.

\subsubsection{Discussion}

\begin{figure}[tb]
    \centering
    \includegraphics[width=\dyncolwidth]{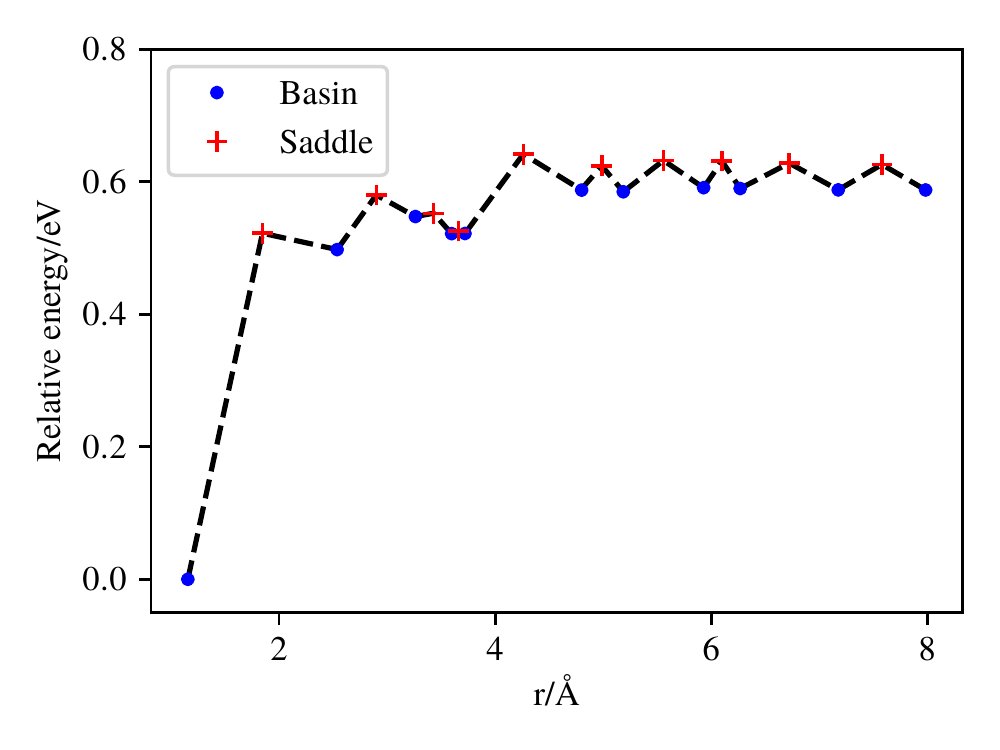}
    \caption{\label{fig:v1h1_meta}Energy profile of the $\text{V}_1\text{H}$ dissociation pathway as a function of the distance of the H atom from the centre of the vacancy.}
\end{figure}

As expected, in \cref{tab:h_energy} we see increasing the number of vacancies leads to a higher $r_x$; $\text{V}_2\text{H}$ has a larger jump in $r_x$ than the rest of the clusters, this is due to extended size of the \num{4}\textsuperscript{th}~NN intermediate state. All of the clusters have a larger effective size than the cluster radius would suggest. This is well demonstrated by $\text{V}_1\text{H}$, where we see $r_x = 3.9$\si{\angstrom} is between the \num{2}\textsuperscript{nd} and \num{3}\textsuperscript{rd}~NN distances, indicating the tetrahedral sites close to the vacancy also act as a weak trap for H. If we examine the energy profile of the dissociation mechanism in \cref{fig:v1h1_meta}, we see a collection of metastable states just below $r = 3.9$\si{\angstrom}. These metastable states join the vacancy's superbasin; as the state-to-state dynamics are not preserved inside superbasins the time fractions at $r < r_x$ in \cref{fig:or_plt} do not correspond to the Boltzmann distribution. Understanding the depth and density of hydrogen's weak-trapping/metastable states introduced by defects is of key importance to designing HE resistant steels that utilise H traps. Our results show that even the simplest defects have a complex secondary-structure of surrounding metastable states. Furthermore, the large effective size of the defects means they will alter the diffusive cross section to a greater extent than would be expected for point defects and such variations are not monotonic with the defect complexity (\ie the number of vacancies).

The detrapping barriers and binding energies in \cref{tab:h_energy} increase with the number of vacancies and the two are separated by $\approx 0.047$\si{\eV} for each complex. Therefore, larger clusters act as deeper traps, this trend should level-out as the inside of the cluster approaches a surface.

The diffusivities of the H atom follow a more complex trend. In general, $D_\text{H}$ and $D_\text{Or}$ fall in accordance with $\Delta E_\text{detrap}$ and $\Delta E_B$ rising but, $D_\text{Or}$ under-predicts the diffusivity. This is because $D_\text{Or}$ assumes the traps are immobile while in reality the energy barrier for the complex diffusing can be very close to the detrapping barrier. This would present a minor contribution to the diffusivity of the H atom if the mean free path (MFP) in the lattice was large, as the diffusion in the lattice is very fast. However due to the high defect-concentration limiting the MFP, the two effects are competing. Furthermore, $D_\text{Or}$ becomes a worse estimator of $D_\text{H}$ as the defect size increases, this is because the approximations made by \citeauthor{Oriani1970} (point trapping, single trapping barrier and no change in diffusive cross-section) become less true as the defect's size increases. The trend is broken by $\text{V}_4\text{H}$, despite having a higher detrapping barrier, the H atom diffuses faster than in $\text{V}_1\text{H}$ and $\text{V}_2\text{H}$. This could be partially explained by fact that $\text{V}_4\text{H}$ has the smallest gap between $\Delta E$ of the complex and $\Delta E_\text{detrap}$ thus, complex diffusion is contributing more to $D_\text{H}$. The corresponding rise in $D_\text{Or}$ (which should not take into account complex diffusion) could then be attributed to the H atom not having enough time to reach equilibrium between the lattice/defect before the defect diffuses again. Nevertheless, the detrapping barriers for $\text{V}_1\text{H}$ and $\text{V}_2\text{H}$ are still lower than the complex-diffusion barrier of $\text{V}_4\text{H}$; perhaps the difference is made up by the kinetic pre-factors of the rate limiting step(s).

Interpolating \cref{tab:h_energy} we would expect the detrapping barrier for  $\text{V}_3\text{H}$ to be \num{0.67}--\num{0.70}\si{\eV}, this is much higher than the diffusion barrier, which explains the lack of detrapping. If we compare the H diffusivities between the complexes, we see that $\text{V}_3\text{H}$ complex-diffusion offers a pathway for H transport that is about an order of magnitude faster at $300$\si{\kelvin} compared to trapping and detrapping from the other complexes. This is partly due to the aforementioned high defect-concentration -- due to the periodic nature of the system -- limiting the MFP of lattice-diffusing H. Nevertheless, this unexpected result could explain elevated H transport in some scenarios. Furthermore, with the energy barrier for H transport via $\text{V}_3\text{H}$ complex-diffusion being so much lower, this could be the predominant H transport mechanism at lower temperatures.

\section{Conclusions}

We have developed and implemented a tolerant norm-based LE classification method for OLKMC, that is invariant under Euclidean-transformations and permutations of atoms. This has enabled the simulation of the Fe-H system into HE timescales at room temperature, predicting features not previously reported using a single modelling framework. This system, with its small interstitials, multi-stage mechanisms, frequent flickering-problems, varied harmonic pre-factors and sensitive energy-barriers, presents many challenges to model with OLKMC. Nevertheless, many of these have been overcome and OLKMC has proved an invaluable and capable tool for the study of these systems with no \textit{a priori} assumptions of the underlying mechanisms. This is extremely promising for the future of modelling H-defect interactions and improving our understanding of HE at the atomic scale. Specifically, we have:
\begin{itemize}
    \item Investigated the diffusion of small (less than six) vacancy-clusters, with and without the addition of H and found evidence that H can increase the diffusivity of larger clusters.
    \item Fully classified the diffusion pathways of these cluster/complexes (energetically and mechanistically) and understood how H changes their diffusion mechanisms.
    \item Simultaneously, obtained the trapping/detrapping barrier(s) of H from -- and its effective diffusivity in -- the presence of these clusters. We have made comparisons to \citeauthor{Oriani1970}'s theory, testing the equilibrium hypothesis in the presence of mobile traps and expanded the conclusions to also include predicting trapping atmospheres in arbitrarily defined non-point traps.
    \item Quantified the trapping atmospheres surrounding vacancy clusters and begun to demonstrate the kinetic effects of shallow traps surrounding point-defects.
    \item Found harmonic pre-factors can vary by up-to two orders-of-magnitude suggesting, the constant pre-factor approximation should always be carefully verified (particularly in multi-element systems).
\end{itemize}
Finally, OLKMC is a materials-agnostic method. Our norm-based classification should find applications in many materials systems requiring long-term atomistic predictions.

\section*{Code availability}

The software used to perform the simulations in this work, including an implementation of the norm-based classification developed, is open-source\footnote{\url{https://github.com/ConorWilliams/onthefly}} or available upon request.

\section*{Acknowledgements}

We gratefully acknowledge the funding received from the EPSRC via the CDT in Computational Methods for Materials Science (grant number EP/L\num{015552}/\num{1}) and grant EP/T\num{008687}/\num{2}.  E. I. Galindo-Nava acknowledges the Royal Academy of Engineering for his research fellowship funding. We also acknowledge Rolls-Royce Plc for the provision of funding. All information and foreground intellectual property generated by this research work is the property of Rolls-Royce Plc.

\bibliographystyle{custom-num-names}

\bibliography{references}

\appendix

\gdef\thesection{\Alph{section}} \makeatletter
\renewcommand\@seccntformat[1]{Appendix \csname the#1\endcsname.\hspace{0.5em}}
\makeatother

\section{\label{app:opp}Orthogonal Procrustes problem}

\begin{algorithm}[H]
    \caption{Solves the orthogonal Procrustes problem, find and return the optimal orthogonal transformation to map the point-cloud $Q$ onto the point-cloud $P$~\cite{Schnemann1966, Zhang2000}.}
    \begin{algorithmic}
        \onehalfspacing
        \Require $P$ and $Q$ contain the same number of points, $n$.
        \Function{rotor\_onto}{$P$, $Q$}
        \State $\bm{H} \gets \sum_{i = 1}^n \bm{q}_i \bm{p}_i\tran$
        \State Compute $\bm{U}$, $\bm{V}$ from the SVD of $\bm{H} $ such that $\bm{H} = \bm{U\Sigma V}\tran$
        \State \Return $\bm{VU}\tran$
        \EndFunction
    \end{algorithmic}
\end{algorithm}

\section{\label{sec:complexity}Complexity analysis of \textsc{greedy\_perm} }

The time complexity of the \textsc{\_recur} function from\cref{alg:permute_onto}, \bigO{R_i}, when called with integer $i$ and point clouds $P$ and $Q$ of size $n$ is:
\begin{align}
    \bigO{R_i} & = \begin{cases}
                       \,(n-i) \left(\bigO{M_i} + \bigO{R_{i+1}}\right) \quad & {i \le n} \\
                       \,n                                                    & {i  > n}
                   \end{cases} \label{equ:complexity}
\end{align}
with \bigO{M_i} the time complexity of \textsc{\_match} called with integer $i$ and point clouds $P$ and $Q$. This is clearly exponential in $n$ for general point clouds. However, in \num{3}D, each intra-point pair in $P$ defines a (thin) shell of thickness $2\sqrt{2}\delta$ around the corresponding points in $Q$, which the next point must fall inside of for \textsc{\_match} to return \texttt{True}. As the distance from four non-coplanar points in \num{3}D define a unique point, once we have matched a small, constant number of points that span the LE, we expect the intersections of these shells to converge to a sphere. Therefore, the volume of space that the tolerance defines in $Q$ is:
\begin{align}
    V \approx \frac{8}{3} \sqrt{2}\pi\delta^3
\end{align}
and point the density, $\rho$, is bound in the worst case by closely-packed spheres of radius $\frac{r_\text{min}}{2}$~\cite{Hales2006}:
\begin{align}
    \rho \le \frac{\sqrt{2}}{r^3_\text{min}}
\end{align}
hence, the expected number of points inside the volume of tolerant space, $m$, is approximately:
\begin{align}
    m \approx \frac{16\pi}{3} \left( \frac{\delta}{r_\text{min}} \right)^3 \label{equ:num_pnt}
\end{align}
Each iteration will on average: call \textsc{\_match} $m$ times for points that will match, each with \bigO{n} runtime; trigger $m$ recursions; call \textsc{\_match} \bigO{n} times for points that will \textbf{not} match, each with constant runtime. Hence, we can rewrite \cref{equ:complexity}:
\begin{align}
    \bigO{R} & =nm + m \left(nm + m \left( \ldots \right) \right) \nonumber \\
    \bigO{R} & = \begin{cases}
                     \,n^2 \quad & m \le 1 \\
                     \,n m^n     & m > 1
                 \end{cases}
\end{align}
In order to avoid exponential complexity, we require $m \le 1$. Rearranging \cref{equ:num_pnt} and using the approximate nature to simplify the constants, this requires:
\begin{align}
    \delta \lesssim \frac{2}{5} r_\text{min} \label{equ:den_app}
\end{align}
For the case of $\alpha$-Fe with $ r_\text{min} \approx 1\si{\angstrom}$, this requires $\delta \lesssim 0.4\si{\angstrom}$. This matches our empirical experiments with typical LEs in $\alpha$-Fe where, we observe the runtime of \cref{alg:permute_onto} blowing-up beyond $\delta \gtrsim 0.4\si{\angstrom}$.

\end{document}


\begin{frontmatter}

    \title{Supplementary material for: Accelerating off-lattice kinetic Monte Carlo simulations to predict hydrogen vacancy-cluster interactions in \texorpdfstring{$\alpha$}{alpha}--Fe}

    \author[cam]{C.J. Williams\corref{cor1}}
    \ead{cw648@cam.ac.uk}

    \affiliation[cam]{
        organization={Department of Materials Science and Metallurgy},
        addressline={University of Cambridge},
        city={Cambridge},
        postcode={CB3 0FS},
        country={UK}
    }

    \cortext[cor1]{Corresponding author}

    \author[cam,ucl]{E.I. Galindo-Nava\corref{cor1}}
    \ead{e.galindo-nava@ucl.ac.uk}

    \affiliation[ucl]{
        organization={ Department of Mechanical Engineering},
        addressline={University College London},
        city={London},
        postcode={WC1E 6BT},
        country={UK}
    }

\end{frontmatter}

\section{Analytical form of the embedded atom method (EAM) Hessian}

\subsection{EAM potential form}

Adopting the notation of Greek superscripts for atom indexes (and later Roman subscripts for vector components) the generalised EAM potential form is~\cite{Ramasubramaniam2009}:
\begin{align}
    U =  \frac{1}{2} \sum_{\substack{\alpha,\mkern1.5mu\beta \\ \alpha \neq \beta}} V^{\alpha\beta}
    + \sum_{\beta} F^{\beta} \label{equ:U}
\end{align}
with $V^{\alpha\beta}$ a symmetric pair-potential function (such that $V^{\alpha\beta} = V^{\beta\alpha}$) acting on $r^{\alpha\beta} = \norm{\bm{r}^{\alpha\beta}} = \norm{\bm{r}^{\beta} - \bm{r}^{\alpha}}$ the atomic separations. Additionally, $F^{\beta}$ is the embedding function acting on $\rho^{\beta}$, the electron density of the $\beta\textsuperscript{th}$ atom~\cite{Ramasubramaniam2009}:
\begin{align}
    \rho^{\beta} & = \sum_{\alpha \neq \beta} \phi^{\alpha\beta} \label{equ:rho}
\end{align}
where $\phi^{\alpha\beta}$ describes the electron density of the $\alpha\textsuperscript{th}$ atom acting at the $\beta\textsuperscript{th}$ atom. Finally, the functional forms of $V$ and $\phi$ are chosen such that:
\begin{align}
    V\left(r^{\alpha\beta}\right)=\phi\left(r^{\alpha\beta}\right)=0
    \quad \text{when} \quad
    r^{\alpha\beta} > r_\text{cut}
\end{align}

\subsection{Partial vector-derivative of the distance between atoms}

Starting from the definition of $r^{\alpha\beta}$, the distance between atoms $\alpha$ and $\beta$:
\begin{align}
    \left[r^{\alpha\beta}\right]^2                                         & =  r^{\alpha\beta}_i r^{\alpha\beta}_i \nonumber                                        \\
    \frac{\partial}{\partial r^{\gamma}_j} \left[r^{\alpha\beta}\right]^2  & = \frac{\partial}{\partial r^{\gamma}_j}  r^{\alpha\beta}_i r^{\alpha\beta}_i \nonumber \\
    r^{\alpha\beta} \frac{\partial r^{\alpha\beta}}{\partial r^{\gamma}_j} & = r^{\alpha\beta}_i \frac{\partial r^{\alpha\beta}_i}{\partial r^{\gamma}_j} \nonumber  \\
    \frac{\partial r^{\alpha\beta}}{\partial r^{\gamma}_j}                 & = \hat{r}^{\alpha\beta}_i \frac{\partial r^{\alpha\beta}_i}{\partial r^{\gamma}_j}
\end{align}
then expanding $r^{\alpha\beta}_i = r^{\beta}_i - r^{\alpha}_i$, with $\bm{r}^{\alpha}$ the position vector of the $\alpha\textsuperscript{th}$ atom:
\begin{align}
    \pd{r^{\alpha\beta}}{r^{\gamma}_j} & = \hat{r}^{\alpha\beta}_i  \left( \frac{\partial r^{\beta}_i}{\partial r^{\gamma}_j} - \frac{\partial r^{\alpha}_i}{\partial r^{\gamma}_j} \right) \nonumber \\
                                       & = \hat{r}^{\alpha\beta}_i  \left( \delta^{\gamma\beta}\delta_{ij} - \delta^{\gamma\alpha}\delta_{ij} \right)                                       \nonumber \\
                                       & = \left( \delta^{\gamma\beta} - \delta^{\gamma\alpha} \right) \hat{r}^{\alpha\beta}_j \label{equ:partial_form}
\end{align}
we arrive at the partial vector-derivative of the distance between atoms.

\subsection{Derivation}

Differentiating \cref{equ:U} and applying \cref{equ:partial_form}, the first derivative of the potential takes the form:
\begin{align}
    \pd{U}{r_j^{\gamma}}
     & = \frac{1}{2} \sum_{\substack{\alpha,\mkern1.5mu\beta                                                                                                                    \\ \alpha \neq \beta}} \dot{V}^{\alpha\beta}  \left( \delta^{\gamma\beta}-\delta^{\gamma\alpha}\right) \hat{r}_j^{\alpha\beta}
    +  \sum_{\beta} \dot{F}^{\beta}\sum_{\alpha \neq \beta} \dot{\phi}^{\alpha\beta} \left( \delta^{\gamma\beta}-\delta^{\gamma\alpha}\right) \hat{r}_j^{\alpha\beta} \nonumber \\
     & =\sum_{\alpha\neq\gamma} \dot{V}^{\alpha\gamma} \hat{r}_j^{\alpha\gamma}
    +  \sum_{\substack{\alpha,\mkern1.5mu\beta                                                                                                                                  \\ \alpha \neq \beta}} \dot{F}^{\beta} \dot{\phi}^{\alpha\beta} \left( \delta^{\gamma\beta}-\delta^{\gamma\alpha}\right) \hat{r}_j^{\alpha\beta} \nonumber \\
     & = \sum_{\alpha\neq\gamma}\left(\dot{V}^{\alpha\gamma} + \dot{F}^{\gamma}\dot{\phi}^{\alpha\gamma}
    + \dot{F}^{\alpha}\dot{\phi}^{\gamma\alpha} \right) \hat{r}_j^{\alpha\gamma}  \label{equ:fall}
\end{align}
where we use an over-dot to denote differentiation and have applied the antisymmetry of $\bm{r}^{\alpha\beta}$. Differentiating \cref{equ:fall} again we obtain:
\begin{align}
    \frac{\partial^2 U}{\partial r_i^{\eta} \partial r_j^{\gamma}}
     & =  \sum_{\alpha\neq\gamma}
    \hat{r}_j^{\alpha\gamma}
    \frac{\partial}{\partial r_i^\eta}
    \left(\dot{V}^{\alpha\gamma} + \dot{F}^{\gamma}\dot{\phi}^{\alpha\gamma} + \dot{F}^{\alpha}\dot{\phi}^{\gamma\alpha} \right) \nonumber                    \\
     & + \sum_{\alpha\neq\gamma} \left(\dot{V}^{\alpha\gamma} + \dot{F}^{\gamma}\dot{\phi}^{\alpha\gamma} + \dot{F}^{\alpha}\dot{\phi}^{\gamma\alpha} \right)
    \frac{\partial \hat{r}_j^{\alpha\gamma}}{\partial r_i^{\eta}} \label{equ:hess_1}
\end{align}
To simplify \cref{equ:hess_1} we begin by separating the first term:
\begin{align}
    \frac{\partial^2 U}{\partial r_i^{\eta} \partial r_j^{\gamma}}
     & = \sum_{\alpha\neq\gamma}
    \hat{r}_j^{\alpha\gamma}
    \frac{\partial  \dot{V}^{\alpha\gamma} }{\partial r_i^\eta}
    \nonumber                                                                                                                                                 \\
     & + \sum_{\alpha\neq\gamma}
    \hat{r}_j^{\alpha\gamma}
    \frac{\partial}{\partial r_i^\eta}
    \left(\dot{F}^{\gamma}\dot{\phi}^{\alpha\gamma} + \dot{F}^{\alpha}\dot{\phi}^{\gamma\alpha} \right) \nonumber                                             \\
     & + \sum_{\alpha\neq\gamma} \left(\dot{V}^{\alpha\gamma} + \dot{F}^{\gamma}\dot{\phi}^{\alpha\gamma} + \dot{F}^{\alpha}\dot{\phi}^{\gamma\alpha} \right)
    \frac{\partial \hat{r}_j^{\alpha\gamma}}{\partial r_i^{\eta}} \label{equ:hess_A1}
\end{align}
then make progress on the partial-derivative in the final line:
\begin{align}
    \pd{\hat{r}_j^{\alpha\gamma}}{r_i^{\eta}} & =  \pd{}{r_i^{\eta}} \frac{r_j^{\alpha\gamma}}{r^{\alpha\gamma}}                                                                                                = \frac{1}{r^{\alpha\gamma}}  \pd{r_j^{\alpha\gamma}}{r_i^{\eta}} - r_j^{\alpha\gamma} \left(\frac{1}{r^{\alpha\gamma}}\right)^2 \pd{r^{\alpha\gamma}}{r^{\eta}_i}
\end{align}
by applying \cref{equ:partial_form} and $r^{\alpha\gamma}_j = r^{\gamma}_j - r^{\alpha}_j$ we find:
\begin{align}
    \pd{\hat{r}_j^{\alpha\gamma}}{r_i^{\eta}} & =  \frac{1}{r^{\alpha\gamma}}  \pd{}{r_i^{\eta}}\left(r^{\gamma}_j - r^{\alpha}_j\right)
    -\left(\frac{1}{r^{\alpha\gamma}}\right)^3 r^{\alpha\gamma}_j r^{\alpha\gamma}_i  \left( \delta^{\eta\gamma} - \delta^{\eta\alpha} \right)                                     \nonumber                      \\
                                              & = \frac{1}{r^{\alpha\gamma}}  \left(\delta^{\eta\gamma} - \delta^{\eta\alpha}\right)\delta_{ij}
    -\frac{1}{r^{\alpha\gamma}} \hat{r}^{\alpha\gamma}_j \hat{r}^{\alpha\gamma}_i  \left(\delta^{\eta\gamma} - \delta^{\eta\alpha}\right)                                        \nonumber                        \\
                                              & = \frac{1}{r^{\alpha\gamma}}  \left(\delta^{\eta\gamma} - \delta^{\eta\alpha}\right)\left(\delta_{ij} - \hat{r}^{\alpha\gamma}_i \hat{r}^{\alpha\gamma}_j \right)
\end{align}
Substituting this back into \cref{equ:hess_A1} and splitting up the second line, we obtain:
\begin{align}
    \frac{\partial^2 U}{\partial r_i^{\eta} \partial r_j^{\gamma}}
     & = \sum_{\alpha\neq\gamma}
    \hat{r}_j^{\alpha\gamma}
    \ddot{V}^{\alpha\gamma} \frac{\partial r^{\alpha\gamma}}{\partial r_i^\eta}\nonumber                                                                                                      \\
     & + \sum_{\alpha\neq\gamma}
    \hat{r}_j^{\alpha\gamma}
    \left(\dot{F}^{\gamma} \frac{\partial \dot{\phi}^{\alpha\gamma} }{\partial r_i^\eta}  + \dot{\phi}^{\alpha\gamma}\frac{\partial \dot{F}^{\gamma} }{\partial r_i^\eta}   \right) \nonumber \\
     & + \sum_{\alpha\neq\gamma}
    \hat{r}_j^{\alpha\gamma}
    \left(\dot{F}^{\alpha} \frac{\partial \dot{\phi}^{\gamma\alpha} }{\partial r_i^\eta}  + \dot{\phi}^{\gamma\alpha}\frac{\partial \dot{F}^{\alpha} }{\partial r_i^\eta}   \right) \nonumber \\
     & + \sum_{\alpha\neq\gamma} \left(\dot{V}^{\alpha\gamma} + \dot{F}^{\gamma}\dot{\phi}^{\alpha\gamma} + \dot{F}^{\alpha}\dot{\phi}^{\gamma\alpha} \right)
    \frac{1}{r^{\alpha\gamma}}\left( \delta^{\eta\gamma}-\delta^{\eta\alpha}\right) \left(\delta_{ij} - \hat{r}_i ^{\alpha\gamma}\hat{r}_j^{\alpha\gamma}\right) \label{equ:hess_A3}
\end{align}
Taking \cref{equ:hess_A3}, we expand the derivatives and contracting the final sum:
\begin{align}
    \frac{\partial^2 U}{\partial r_i^{\eta} \partial r_j^{\gamma}}
     & = \sum_{\alpha\neq\gamma}
    \hat{r}_j^{\alpha\gamma}
    \ddot{V}^{\alpha\gamma} \left( \delta^{\eta\gamma}-\delta^{\eta\alpha}\right) \hat{r}_i^{\alpha\gamma}\nonumber                                                                                                            \\
     & + \sum_{\alpha\neq\gamma}
    \hat{r}_j^{\alpha\gamma}
    \left(\dot{F}^{\gamma} \ddot{\phi}^{\alpha\gamma} \frac{\partial r^{\alpha\gamma} }{\partial r_i^\eta}  + \dot{\phi}^{\alpha\gamma} \ddot{F}^{\gamma} \frac{\partial \rho^{\gamma}}{\partial r_i^\eta}   \right) \nonumber \\
     & + \sum_{\alpha\neq\gamma}
    \hat{r}_j^{\alpha\gamma}
    \left(\dot{F}^{\alpha} \ddot{\phi}^{\gamma\alpha} \frac{\partial r^{\gamma\alpha} }{\partial r_i^\eta}  + \dot{\phi}^{\gamma\alpha} \ddot{F}^{\alpha} \frac{\partial \rho^{\alpha}}{\partial r_i^\eta}   \right) \nonumber \\
     & + \delta^{\eta\gamma} \sum_{\alpha\neq\gamma} \left(\dot{V}^{\alpha\gamma} + \dot{F}^{\gamma}\dot{\phi}^{\alpha\gamma} + \dot{F}^{\alpha}\dot{\phi}^{\gamma\alpha} \right)
    \frac{1}{r^{\alpha\gamma}} \left(\delta_{ij} - \hat{r}_i ^{\alpha\gamma}\hat{r}_j^{\alpha\gamma}\right)            \nonumber                                                                                               \\
     & - \left(1 - \delta^{\eta\gamma}\right)\left(\dot{V}^{\eta\gamma} + \dot{F}^{\gamma}\dot{\phi}^{\eta\gamma} + \dot{F}^{\eta}\dot{\phi}^{\gamma\eta} \right)
    \frac{1}{r^{\eta\gamma}}\left(\delta_{ij} - \hat{r}_i ^{\eta\gamma}\hat{r}_j^{\eta\gamma}\right)
\end{align}
then contracting the first sum and expanding the derivatives a second time to arrive at:
\begin{align}
    \frac{\partial^2 U}{\partial r_i^{\eta} \partial r_j^{\gamma}}
     & = \delta^{\eta\gamma} \sum_{\alpha\neq\gamma}
    \ddot{V}^{\alpha\gamma} \hat{r}_i^{\alpha\gamma}\hat{r}_j^{\alpha\gamma} - \left(1  - \delta^{\eta\gamma}\right) \ddot{V}^{\eta\gamma} \hat{r}_i^{\eta\gamma}\hat{r}_j^{\eta\gamma} \nonumber                                                                                                                                   \\
     & + \sum_{\alpha\neq\gamma}
    \hat{r}_j^{\alpha\gamma}
    \left(\dot{F}^{\gamma} \ddot{\phi}^{\alpha\gamma} \left( \delta^{\eta\gamma}-\delta^{\eta\alpha}\right) \hat{r}_i^{\alpha\gamma}  + \dot{\phi}^{\alpha\gamma} \ddot{F}^{\gamma} \sum_{\beta \neq \gamma} \dot{\phi}^{\beta\gamma} \left( \delta^{\eta\gamma}-\delta^{\eta\beta}\right) \hat{r}_i^{\beta\gamma}\right) \nonumber \\
     & + \sum_{\alpha\neq\gamma}
    \hat{r}_j^{\alpha\gamma}
    \left(\dot{F}^{\alpha} \ddot{\phi}^{\gamma\alpha} \left( \delta^{\eta\alpha}-\delta^{\eta\gamma}\right) \hat{r}_i^{\gamma\alpha}  + \dot{\phi}^{\gamma\alpha} \ddot{F}^{\alpha} \sum_{\beta \neq \alpha} \dot{\phi}^{\beta\alpha} \left( \delta^{\eta\alpha}-\delta^{\eta\beta}\right) \hat{r}_i^{\beta\alpha}\right) \nonumber \\
     & + \delta^{\eta\gamma} \sum_{\alpha\neq\gamma} \left(\dot{V}^{\alpha\gamma} + \dot{F}^{\gamma}\dot{\phi}^{\alpha\gamma} + \dot{F}^{\alpha}\dot{\phi}^{\gamma\alpha} \right)
    \frac{1}{r^{\alpha\gamma}} \left(\delta_{ij} - \hat{r}_i ^{\alpha\gamma}\hat{r}_j^{\alpha\gamma}\right)            \nonumber                                                                                                                                                                                                    \\
     & - \left(1 - \delta^{\eta\gamma}\right)\left(\dot{V}^{\eta\gamma} + \dot{F}^{\gamma}\dot{\phi}^{\eta\gamma} + \dot{F}^{\eta}\dot{\phi}^{\gamma\eta} \right)
    \frac{1}{r^{\eta\gamma}}\left(\delta_{ij} - \hat{r}_i ^{\eta\gamma}\hat{r}_j^{\eta\gamma}\right)
\end{align}
wherein we split up the second and third lines before recombining them to yield:
\begin{align}
    \frac{\partial^2 U}{\partial r_i^{\eta} \partial r_j^{\gamma}}
     & = \delta^{\eta\gamma} \sum_{\alpha\neq\gamma}
    \ddot{V}^{\alpha\gamma} \hat{r}_i^{\alpha\gamma}\hat{r}_j^{\alpha\gamma} - \left(1  - \delta^{\eta\gamma}\right) \ddot{V}^{\eta\gamma} \hat{r}_i^{\eta\gamma}\hat{r}_j^{\eta\gamma} \nonumber \\
     & + \sum_{\alpha\neq\gamma}
    \hat{r}_j^{\alpha\gamma}
    \dot{F}^{\gamma} \ddot{\phi}^{\alpha\gamma} \left( \delta^{\eta\gamma}-\delta^{\eta\alpha}\right) \hat{r}_i^{\alpha\gamma}
    +  \hat{r}_j^{\alpha\gamma} \dot{F}^{\alpha} \ddot{\phi}^{\gamma\alpha} \left( \delta^{\eta\alpha}-\delta^{\eta\gamma}\right) \hat{r}_i^{\gamma\alpha}\nonumber                               \\
     & + \sum_{\alpha\neq\gamma} \sum_{\beta \neq \gamma}
    \hat{r}_j^{\alpha\gamma}
    \dot{\phi}^{\alpha\gamma} \ddot{F}^{\gamma}  \dot{\phi}^{\beta\gamma} \left( \delta^{\eta\gamma}\right) \hat{r}_i^{\beta\gamma}
    \nonumber                                                                                                                                                                                     \\
     & + \sum_{\alpha\neq\gamma} \sum_{\beta \neq \alpha}
    \hat{r}_j^{\alpha\gamma}
    \dot{\phi}^{\gamma\alpha} \ddot{F}^{\alpha}  \dot{\phi}^{\beta\alpha} \left( \delta^{\eta\alpha}\right) \hat{r}_i^{\beta\alpha} - \sum_{\alpha\neq\gamma} \sum_{\beta \neq \gamma}
    \hat{r}_j^{\alpha\gamma}
    \dot{\phi}^{\alpha\gamma} \ddot{F}^{\gamma}  \dot{\phi}^{\beta\gamma} \left(\delta^{\eta\beta}\right) \hat{r}_i^{\beta\gamma}
    \nonumber                                                                                                                                                                                     \\
     & - \sum_{\alpha\neq\gamma} \sum_{\beta \neq \alpha}
    \hat{r}_j^{\alpha\gamma}
    \dot{\phi}^{\gamma\alpha} \ddot{F}^{\alpha}  \dot{\phi}^{\beta\alpha} \left(\delta^{\eta\beta}\right) \hat{r}_i^{\beta\alpha} \nonumber                                                       \\
     & + \delta^{\eta\gamma} \sum_{\alpha\neq\gamma} \left(\dot{V}^{\alpha\gamma} + \dot{F}^{\gamma}\dot{\phi}^{\alpha\gamma} + \dot{F}^{\alpha}\dot{\phi}^{\gamma\alpha} \right)
    \frac{1}{r^{\alpha\gamma}} \left(\delta_{ij} - \hat{r}_i ^{\alpha\gamma}\hat{r}_j^{\alpha\gamma}\right)            \nonumber                                                                  \\
     & - \left(1 - \delta^{\eta\gamma}\right)\left(\dot{V}^{\eta\gamma} + \dot{F}^{\gamma}\dot{\phi}^{\eta\gamma} + \dot{F}^{\eta}\dot{\phi}^{\gamma\eta} \right)
    \frac{1}{r^{\eta\gamma}}\left(\delta_{ij} - \hat{r}_i ^{\eta\gamma}\hat{r}_j^{\eta\gamma}\right)
\end{align}
finally, contracting the remaining Kronecker deltas, we arrive at an expression for the Hessian:
\begin{align}
    \frac{\partial^2 U}{\partial r_i^{\eta} \partial r_j^{\gamma}}
     & = \delta^{\eta\gamma} \sum_{\alpha\neq\gamma}
    \ddot{V}^{\alpha\gamma} \hat{r}_i^{\alpha\gamma}\hat{r}_j^{\alpha\gamma} - \left(1  - \delta^{\eta\gamma}\right) \ddot{V}^{\eta\gamma} \hat{r}_i^{\eta\gamma}\hat{r}_j^{\eta\gamma} \nonumber \\
     & + \delta^{\eta\gamma} \sum_{\alpha\neq\gamma}
    \left(\dot{F}^{\gamma} \ddot{\phi}^{\alpha\gamma}
    + \dot{F}^{\alpha} \ddot{\phi}^{\gamma\alpha} \right) \hat{r}_i^{\alpha\gamma}\hat{r}_j^{\alpha\gamma} - \left(1  - \delta^{\eta\gamma}\right)
    \left(\dot{F}^{\gamma} \ddot{\phi}^{\eta\gamma}
    +\dot{F}^{\eta} \ddot{\phi}^{\gamma\eta}
    \right) \hat{r}_i^{\eta\gamma}\hat{r}_j^{\eta\gamma}   \nonumber                                                                                                                              \\
     & + \delta^{\eta\gamma} \sum_{\alpha\neq\gamma} \sum_{\beta \neq \gamma}
    \hat{r}_j^{\alpha\gamma}
    \dot{\phi}^{\alpha\gamma} \ddot{F}^{\gamma}  \dot{\phi}^{\beta\gamma} \hat{r}_i^{\beta\gamma} \nonumber                                                                                       \\
     & + \left(1 - \delta^{\eta\gamma}\right) \left[ \sum_{\alpha \neq \eta}
    \hat{r}_j^{\eta\gamma}
    \dot{\phi}^{\gamma\eta} \ddot{F}^{\eta}  \dot{\phi}^{\alpha\eta}  \hat{r}_i^{\alpha\eta} - \sum_{\alpha\neq\gamma}
    \hat{r}_j^{\alpha\gamma}
    \dot{\phi}^{\alpha\gamma} \ddot{F}^{\gamma}  \dot{\phi}^{\eta\gamma} \hat{r}_i^{\eta\gamma}
    \right] \nonumber                                                                                                                                                                             \\
     & -\sum_{\alpha\neq\gamma,\, \eta}
    \hat{r}_j^{\alpha\gamma}
    \dot{\phi}^{\gamma\alpha} \ddot{F}^{\alpha}  \dot{\phi}^{\eta\alpha}\hat{r}_i^{\eta\alpha} \nonumber                                                                                          \\
     & + \delta^{\eta\gamma} \sum_{\alpha\neq\gamma} \left(\dot{V}^{\alpha\gamma} + \dot{F}^{\gamma}\dot{\phi}^{\alpha\gamma} + \dot{F}^{\alpha}\dot{\phi}^{\gamma\alpha} \right)
    \frac{1}{r^{\alpha\gamma}} \left(\delta_{ij} - \hat{r}_i ^{\alpha\gamma}\hat{r}_j^{\alpha\gamma}\right)            \nonumber                                                                  \\
     & - \left(1 - \delta^{\eta\gamma}\right)\left(\dot{V}^{\eta\gamma} + \dot{F}^{\gamma}\dot{\phi}^{\eta\gamma} + \dot{F}^{\eta}\dot{\phi}^{\gamma\eta} \right)
    \frac{1}{r^{\eta\gamma}}\left(\delta_{ij} - \hat{r}_i ^{\eta\gamma}\hat{r}_j^{\eta\gamma}\right) \label{equ:hess_A2}
\end{align}

\subsection{Simplification}

\Cref{equ:hess_A2} has two very-separate regimes: the block-diagonal, for which $\eta = \gamma$ and the off-diagonal elements, for which $\eta \ne \gamma$. Exploring the block-diagonal elements:
\begin{align}
    \frac{\partial^2 U}{\partial r_i^{\gamma} \partial r_j^{\gamma}}
     & = \sum_{\alpha\neq\gamma}
    \left(\ddot{V}^{\alpha\gamma} + \dot{F}^{\gamma} \ddot{\phi}^{\alpha\gamma}
    + \dot{F}^{\alpha} \ddot{\phi}^{\gamma\alpha} + \ddot{F}^{\alpha} \dot{\phi}^{\gamma\alpha} \dot{\phi}^{\gamma\alpha} \right) \hat{r}_i^{\alpha\gamma}\hat{r}_j^{\alpha\gamma} \nonumber
    \\
     & +\ddot{F}^{\gamma} \left[\sum_{\beta \neq \gamma}
    \dot{\phi}^{\beta\gamma} \hat{r}_i^{\beta\gamma}\right] \left[\sum_{\alpha \neq \gamma}
    \dot{\phi}^{\alpha\gamma} \hat{r}_j^{\alpha\gamma}\right]
    \nonumber                                                                                                                                                 \\
     & + \sum_{\alpha\neq\gamma} \left(\dot{V}^{\alpha\gamma} + \dot{F}^{\gamma}\dot{\phi}^{\alpha\gamma} + \dot{F}^{\alpha}\dot{\phi}^{\gamma\alpha} \right)
    \frac{1}{r^{\alpha\gamma}} \left(\delta_{ij} - \hat{r}_i ^{\alpha\gamma}\hat{r}_j^{\alpha\gamma}\right)
\end{align}
which, if we introduce the electron-density dipole:
\begin{align}
    \mu_i^\beta = \sum_{\alpha \neq \beta} \dot{\phi}^{\alpha\beta} \hat{r}_i^{\alpha\beta}
\end{align}
and symmetric tensors $A$, $B$ such that:
\begin{align}
    A^{\alpha\beta}  = \frac{\dot{V}^{\alpha\beta} + \dot{F}^{\beta}\dot{\phi}^{\alpha\beta} + \dot{F}^{\alpha}\dot{\phi}^{\beta\alpha}}{r^{\alpha\beta}}
    \quad \text{and} \quad
    B^{\alpha\beta}  = \ddot{V}^{\alpha\gamma} + \dot{F}^{\gamma} \ddot{\phi}^{\alpha\gamma} + \dot{F}^{\alpha} \ddot{\phi}^{\gamma\alpha}
\end{align}
becomes:
\begin{align}
    \boxed{\frac{\partial^2 U}{\partial r_i^{\gamma} \partial r_j^{\gamma}} = \ddot{F}^{\gamma} \mu_i^\gamma \mu_j^\gamma + \sum_{\alpha\neq\gamma} A^{\alpha\gamma} \delta_{ij} +  \left(A^{\alpha\gamma} - B^{\alpha\gamma}
    - \ddot{F}^{\alpha} \dot{\phi}^{\gamma\alpha} \dot{\phi}^{\gamma\alpha}
    \right) \hat{r}_i ^{\alpha\gamma}\hat{r}_j^{\gamma\alpha}}
\end{align}
For the off-diagonal elements of the Hessian we can simplify \cref{equ:hess_A2} to:
\begin{align}
    \left.\frac{\partial^2 U}{\partial r_i^{\eta} \partial r_j^{\gamma}}\right|_{\eta \neq \gamma}
     & = - \left(\ddot{V}^{\eta\gamma} + \dot{F}^{\gamma} \ddot{\phi}^{\eta\gamma}
    + \dot{F}^{\eta} \ddot{\phi}^{\gamma\eta}
    \right) \hat{r}_i^{\eta\gamma}\hat{r}_j^{\eta\gamma}\nonumber                                                                  \\
     & + \sum_{\alpha \neq \eta}
    \hat{r}_j^{\eta\gamma}
    \dot{\phi}^{\gamma\eta} \ddot{F}^{\eta}  \dot{\phi}^{\alpha\eta}  \hat{r}_i^{\alpha\eta} - \sum_{\alpha\neq\gamma}
    \hat{r}_j^{\alpha\gamma}
    \dot{\phi}^{\alpha\gamma} \ddot{F}^{\gamma}  \dot{\phi}^{\eta\gamma} \hat{r}_i^{\eta\gamma}
    \nonumber                                                                                                                      \\
     & - \sum_{\alpha\neq\gamma,\, \eta}
    \dot{\phi}^{\gamma\alpha} \ddot{F}^{\alpha}  \dot{\phi}^{\eta\alpha}\hat{r}_i^{\eta\alpha}  \hat{r}_j^{\alpha\gamma} \nonumber \\
     & - \left(\dot{V}^{\eta\gamma} + \dot{F}^{\gamma}\dot{\phi}^{\eta\gamma} + \dot{F}^{\eta}\dot{\phi}^{\gamma\eta} \right)
    \frac{1}{r^{\eta\gamma}}\left(\delta_{ij} - \hat{r}_i ^{\eta\gamma}\hat{r}_j^{\eta\gamma}\right)
\end{align}
which -- after substituting in the gradient-component function and electron-density dipole -- reduces to:
\begin{align}
    \boxed{\left.\frac{\partial^2 U}{\partial r_i^{\eta} \partial r_j^{\gamma}}\right|_{\eta \neq \gamma}
    =
    \left(B^{\eta\gamma} - A^{\eta\gamma}\right) \hat{r}_i^{\eta\gamma}\hat{r}_j^{\gamma\eta} - A^{\eta\gamma}\delta_{ij}
    + \ddot{F}^{\eta} \dot{\phi}^{\gamma\eta} \mu_i^\eta \hat{r}_j^{\eta\gamma}
    - \ddot{F}^{\gamma} \dot{\phi}^{\eta\gamma} \hat{r}_i^{\eta\gamma} \mu_j^\gamma
    -   O^{\eta\gamma}_{ij}}
\end{align}
Here the first four terms are non-zero when $r^{\eta\gamma} < r_\text{cut}$ however, the overlap term:
\begin{align}
    O^{\eta\gamma}_{ij} = \sum_{\alpha\neq\gamma,\, \eta}  \ddot{F}^{\alpha} \dot{\phi}^{\gamma\alpha} \dot{\phi}^{\eta\alpha}\hat{r}_i^{\eta\alpha}  \hat{r}_j^{\alpha\gamma}
\end{align}
is more complex as the sum runs-over the intersection of the neighbours of atoms $\eta$ and $\gamma$ thus, can be non-zero when $r^{\eta\gamma} < 2r_\text{cut}$.

\bibliographystyle{custom-num-names}
\bibliography{references}